\documentclass[aps,prd,reprint,superscriptaddress,nofootinbib]{revtex4-1}
\usepackage{cuted} % provides widetext for wide equations in two-column mode
\usepackage[normalem]{ulem}
\usepackage[normalem]{ulem}
\usepackage{xcolor}
\usepackage{mdframed}

\usepackage[T1]{fontenc} % if needed

\usepackage{graphicx,epsfig}% Include figure files
\usepackage{bbm}
\usepackage{graphics,dcolumn,bm, float}
\usepackage{amssymb,amsmath,rotate,color,amsfonts}
\usepackage[title,titletoc,toc]{appendix}
\usepackage{mathtools}
\usepackage{booktabs}
\usepackage{tcolorbox}
\usepackage{tikz-cd}
\usetikzlibrary{shadings}
\usepackage{xcolor}
\usepackage{tikz-3dplot}
\usepackage{setspace}
\usepackage{wrapfig}
\usepackage{lipsum}
\usepackage{mwe}
\usepackage[mathlines]{lineno}
\usepackage[mathscr]{euscript}
\usepackage{bbold}
\usepackage{pgfplots}
\usepackage{float}
\usepackage{tkz-euclide}
\usepackage{braket}
\usepackage{physics}
\usepackage[caption=false]{subfig}
\usepackage[export]{adjustbox}
\usepackage{tikz}
\usepackage{slashed}
\usepackage{bm}
\usepackage{multirow}
\usetikzlibrary{through,calc}
\usetikzlibrary{positioning}

\usepackage{hyperref}

\newcommand{\be}{\begin{equation}}
	\newcommand{\ee}{\end{equation}}
\newcommand{\bea}{\begin{eqnarray}}
	\newcommand{\eea}{\end{eqnarray}}
\newcommand{\nn}{\nonumber}
\newcommand{\normord}[1]{:\mathrel{#1}:}
\newcolumntype{P}[1]{>{\centering\arraybackslash}p{#1}}
\tdplotsetmaincoords{70}{120} % adjust viewing angle

\begin{document}

\nocite{*}

\title{The $Z_N$ equivariant Virasoro algebra via alternative Sugawara constructions}

\preprint{}

\author{Armin Ghazi}
\author{Ahmad Moradpouri}
\affiliation{Research Center for High Energy Physics, Department of Physics, Sharif University of Technology, P.O.Box 11155-9161, Tehran, Iran.}
\thanks{armin.gh.kh@gmail.com;
Ahmadreza.Moradpour@gmail.com}
\begin{abstract}
	In this paper, we study the $u(1)^2$ Kac--Moody algebra and generalize the standard Sugawara construction of the Virasoro algebra to an infinite family of new realizations. In this case, in addition to the standard invariant tensor $\delta^{ij}$, there exists another invariant tensor $\epsilon^{ij}$, which enables 
the construction of genuinely new realizations beyond the conventional one. We show that these new realizations arise from a $\mathbb{Z}_N$-grading of the mode index $n$ of the Virasoro generators $L_n$ and the space of such realizations corresponds to points of a possibly singular algebraic variety. For the $\mathbb{Z}_2$ and $\mathbb{Z}_3$ cases, the space of all such constructions is topologically equivalent to a cylinder, while for $\mathbb{Z}_4$ it forms a non-compact real four-dimensional manifold. We show that the spaces of constructions for $Z_{2N}$ and $Z_{2N+1}$ are closely similar. Furthermore, we reformulate the problem within an action-principle framework by introducing $\mathbb{Z}_N$-equivariant maps, which provide a systematic method for constructing conformal field theories endowed with these generalized Virasoro symmetries. This formulation reproduces the $\mathbb{Z}_2$ case and supports the idea that $\mathbb{Z}_N$-equivariance offers a consistent and unified approach to generating extended conformal algebras. Finally, we analyze the corresponding Virasoro-Kac-Moody-like algebras associated with these constructions and show that they represent nontrivial (up to local field transformations) of the well-known Virasoro-Kac-Moody algebra.
\end{abstract}

	\maketitle
	\flushbottom
	
	%\input{sections/preprocessors.tex}
	%\input{sections/intro.tex}
	%\input{sections/main.tex}
	
	%\newpage
	%\appendix
	%\input{sections/appendix.tex}
	%\input{sections/appendix2.tex}
	
	%\newpage
	%\bibliographystyle{JHEP}
	%\bibliography{mybib}
	
	\section{Introduction}
	Conformal field theory (CFT) in two dimensions has been a cornerstone of theoretical physics for the past several decades. Since the seminal work of Belavin, Polyakov, and Zamolodchikov (BPZ) \cite{BELAVIN1984333}, two-dimensional CFT has provided a powerful and elegant framework for the study of critical phenomena and exactly solvable models~\cite{DiFrancesco:1997nk,ginsparg1988appliedconformalfieldtheory,Blumenhagen:2009zz,Friedan1986NotesOS,Polyakov:1970xd,Friedan:1984rv}. Its applications range from statistical mechanics~\cite{CARDY1986186,cardy2008conformalfieldtheorystatistical,Itzykson:1989sy}, condensed matter physics from Ising model to fractional quantum Hall effect~\cite{tong2016lecturesquantumhalleffect,fradkin2021quantum,MOORE1991362,Fateev:1985mm} to string theory\cite{Polchinski:1998rq,Becker_Becker_Schwarz_2006,Blumenhagen:2013fgp}. 
	It also continues to play an important role in holography. The seminal work of~\cite{brown1986central} showed that the asymptotic symmetry algebra of $AdS_3$ is formed by two copies of the Virasoro algebra. 
	Needless to say, two-dimensional CFTs have a significant and broad contribution to the field of mathematical physics in the last three decades~\cite{VERLINDE1988360,Moore:1988qv,Moore:1989yh,frenkel2005lectureslanglandsprogramconformal,Goddard:1986bp}. Moreover, 2d-CFTs also have many applications in pure mathematics, through intercorrelated concepts. A few examples are vertex operator algebras~\cite{RichardBorcherds,borcherds1999quantumvertexalgebras,Frenkel1988xz}, representation theory of affine Lie algebras~\cite{kac1990infinite,MOODY1968211,Lepowsky:1978jk,frenkel1992vertex,Frenkel:1980rn,Goddard:1986ee} and even using the WZW/Chern-Simon correspondence to knot theory~\cite{Witten:1988hf}.   Many families of 2d-CFTs have been studied by different methods. The minimal Models, which are built out of finitely many Irreducible representations of the Virasoro algebra ~\cite{kac1987_bombay,FeiginFuchs1982} and orbifold CFTs~\cite{Dixon:1985jw,DixonFriedanMartinecShenker1987,DijkgraafVafaVerlindeVerlinde1989} are examples of such families. 
	
	Furthermore, there exist many conformal field theories in which the Virasoro algebra is generated not only by the energy–momentum tensor, but also supplemented by additional symmetries encoded in other fields of the theory. These fields furnish operator product expansions that close under a larger algebraic structure beyond the Virasoro algebra itself. The most prominent and well-studied examples of such extended symmetries are provided by affine Kac–-Moody algebras~\cite{kac1990infinite,MOODY1968211,kac1987_bombay}, which arise as current algebras associated with conserved holomorphic and antiholomorphic currents~\cite{Knizhnik:1984uv}. The Kac-Moody algebra $\hat g_k$ for Lie algebra $g$ and level $k$ is given by
	\bea
	[J^a_m, J^b_n] = i f^{ab}_{\;\;c} J^c_{m+n} + k\, m \, \delta^{ab} \delta_{m+n,0}.
	\eea
	Such symmetry algebras appear in a wide range of contexts. For instance, the affine algebra $\hat{su}(2)_1$ corresponds to a free boson, while more generally, 
	they emerge in non-abelian bosonization~\cite{Witten:1983ar}, in the compactifications of string theory on group manifolds~\cite{GEPNER1986493}, and in the study of various statistical models. Notably, interesting models such as the Potts and Ising models can be realized through coset constructions of the Kac–-Moody algebras~\cite{Goddard:1986bp,Blumenhagen:2009zz}. The symmetry algebra generated by conserved currents, the Kac-Moody algebra, is not independent of the Virasoro algebra. Instead, consistency between these two structures is ensured through the Sugawara construction~\cite{PhysRev.170.1659}, where the energy–momentum tensor is expressed as a quadratic combination of the Kac-Moody currents.

The standard Sugawara construction has been extended in several directions. The general Virasoro affine construction on an affine Lie algebra 
$\mathfrak{g}$ was studied by M.B. Halpern and E. Kritsis~\cite{Halpern:1989ss}, and further explored (and extended) in subsequent works~\cite{HALPERN1991333,HALPERN1990411,Halpern:1989zy,de_Boer_1997}. Such generalizations have also been shown to be connected to problems in quantum mechanics~\cite{morozov}. In this paper, we show that there are many more new realizations for special Kac-Moody algebras.  For the group $g = U(1)^D$, the Kac-Moody algebra is simplified due to the abelian nature of the group. In this case, the algebra reduces to a set of commuting current algebras(up to central term), which is essentially identical to the oscillator algebra satisfied by the $\alpha^i_n$ modes in the free string theory. These $\alpha^i_n$ are the Fourier coefficients of the currents $J^i(z) = \partial X^i(z)$ arising from the mode expansion of free bosons, thereby providing a direct realization of an abelian Kac–Moody algebra within string theory and the Virasoro generators take the usual representations in terms of $\alpha^i_n$ as follows
	\bea
	\label{usualcons}
	L_n=\frac{1}{2}\sum_q \alpha^i_{n-q}\alpha^i_q,
	\eea
	This is consistent with the Sugawara construction. This paper aims to show that such a construction is not unique for $U(1)^2$, and there are infinitely many alternative constructions of $L_n$ in terms of $\alpha^i_n$ that reproduce the Virasoro algebra. Our construction relies on two main ingredients. First, the Virasoro generators can be decomposed into $N$ classes by considering the integer index $n$ modulo $N$. In this framework, the operators are denoted as $L^k_n$~\footnote{In the rest of this paper, we denote $L_n^k=S^k_n$ for $k>0$ and $L^0_n=L_n$.}, where the index takes values of the form $n = N\mathbb{Z} + k$ with $k = 0,1,\dots, N-1$, which is the $\mathbb{Z}_N$ partitioning of the integer numbers module $N$. Consequently, the usual Virasoro generator $L_n$ is naturally decomposed into $N$ distinct classes. Furthermore, since the Hamiltonian of a two-dimensional conformal field theory is given by $H = L_0 + \tilde{L}_0$, we fix the $k=0$ class to correspond to the standard Virasoro construction~\eqref{usualcons}, where $n\in N\mathbb{Z}$, which is Hermitian and positive definite.~\footnote{From a more algebraic perspective, this condition can be relaxed. For instance, in the case $N=2$, the most general form of the operator $S^0_n$ can be written as $L_n=\frac{1}{2}\sum_q(A+B(-1)^q)\alpha^i_{n-q}\alpha^i_q$. However, in the remainder of the paper, we fix $S^0_n$ to be the standard Virasoro operator, postponing the discussion of these more general variants to the section~\ref{actionZ2} where such modifications are considered.} Second, in two dimensions, in addition to the invariant tensor $\delta^{ij}$, there also exists the antisymmetric tensor $\epsilon^{ij}$, which is invariant under rotations. The combination of $\delta^{ij}$ and $\epsilon^{ij}$ is sufficient to generate infinitely many additional constructions of Virasoro-like operators.  
    
     The $\mathbb{Z}_N$ equivariant Virasoro algebra and its constructions in terms of current modes also have an interesting relation to the deformation theory of $u(1)^2$ Kac-Moody-Virasoro algebras. Deformation theory plays a significant role in modern physics, and many examples can be understood from the viewpoint of deformation theory. In quantum physics, for example, deformations are central to the deformation quantization program, where quantum mechanics is viewed as a deformed version of classical Poisson algebra which originated from physical motivations~\cite{Moyal:1949sk,Groenewold:1946kp}(review from physical point of view~\cite{curtright2013concise}) and is one of the cornerstone of mathematical physics with work of many mathematicians specially Kontsevich~\cite{Kontsevich_2003}(for a review from mathematical point of view~\cite{Moshayedi:2022dkl}). Another example is the deformation of Galilean algebra to Poincaré algebra, where the speed of light is the parameter of the deformation~\cite{Figueroa-OFarrill:1989wmj} and Poincaré algebra itself can be deformed to de Sitter algebra~\cite{levy1967deformation}. We should note that, although such constructions in terms of $\alpha^i_{n}$ reproduce the Virasoro algebra, they lead to the deformation of the standard $u(1)^2$ Kac-Moody-Virasoro algebra, which in its own right is interesting.

	The organization of this paper is as follows: In Section~\ref{warmup}, we present a motivating example that illustrates a new construction of the Virasoro algebra in terms of the $\alpha^i_n$ modes, thereby setting the stage for a more detailed investigation of the $\mathbb{Z}_N$ case. In Section~\ref{ModN}, we carry out a systematic study of the new construction and demonstrate that it exhibits a natural $\mathbb{Z}_2$-grading of the integers labeling the Virasoro generators $L_n$ with $n\in\mathbb{Z}$. We then extend the analysis to the general $\mathbb{Z}_N$ case and derive the conditions that these new constructions must satisfy to be consistent with the Virasoro algebra. It turns out that the $\mathbb{Z}_2$ and $\mathbb{Z}_3$ cases admit infinitely many new constructions. The corresponding space of solutions is topologically equivalent to a cylinder, which can be compactified to smooth $\mathbb{CP}^1$ by adjoining two points at infinity, which is a Riemann surface of genus $g=0$.The algebras corresponding to the points at infinity are explored in~\cite{ghazi2026bms3likealgebraszngradedu12}, and it is shown that they are related to the generalization of the $BMS_3$ algebra. We then investigate the structure of the $\mathbb{Z}_4$ case, where the space of new Virasoro generators expressed in terms of $\alpha^i_n$ is four-dimensional. In Section~\ref{equivariantsection}, we approached the new constructions from the perspective of an action principle rather than purely algebraic considerations. In particular, we developed the concept of $\mathbb{Z}_N$-equivariant maps as a systematic method to construct new conformal field theories. Using this framework, one can successfully generate the $\mathbb{Z}_2$ case, providing strong evidence that the notion of $\mathbb{Z}_N$-equivariance is sufficient to produce these algebras in a controlled and consistent manner. In Section~\ref{Deformedalgebras}, we have explored the Virasoro--Kac--Moody-like algebras associated to such constructions, and interestingly, they exhibit interesting deformations of such algebras (up to local field transformations).

	\section{Warm up: A new representation of the Virasoro algebra in d=2}
	\label{warmup}
	As mentioned in the introduction, the connection between the abelian $u(1)^D$ algebra and the Virasoro algebra is not limited to the well-known construction, $L_n=\frac{1}{2}\sum_{q=-\infty}^{\infty}\normord{\alpha^i_{n-q}\alpha^i_q}$ where $\normord{}$ denotes normal ordering. This section aims to explore this relationship in greater depth.
	We claim that, in $D=2$, there exist many additional constructions that also satisfy the Virasoro algebra.
	In two dimensions, besides $\delta^{ij}$, there is also a rotationally invariant tensor $\epsilon^{ij}$, which allows us to define a richer class of operators beyond the standard Virasoro generators.

	As a warm-up, let us define the following operator:
	\bea
	\label{defS_n}
	S_n=\frac{1}{2}\sum_q(-1)^q\epsilon^{ij}\alpha^i_{n-q}\alpha^j_q,
	\eea   
	This is a new operator, well-defined for odd values of $n$ (for even $n$, it simply vanishes).
	The commutators of $S_n$ with $\alpha^i_m$, namely $
	[S_n,\alpha^i_m]
	$, can be computed straightforwardly.
	Since the commutators of $\alpha^i_n$ are $c$-numbers, we may safely ignore normal ordering.
	Consequently, the above commutators simplify to the following:
	\bea
	&&[S_n,\alpha^i_m]=(-1)^mm\epsilon^{ij}\alpha^j_{n+m}.
	\eea
	
	It is straightforward to verify that the set consisting of the standard $L_n$ with even $n$ together with the new operators $S_n$ with odd $n$ satisfies the Virasoro algebra, thus providing an extended realization of the Virasoro algebra beyond the standard construction.
	\bea
	\label{Vir1}
	&&[L_n, L_m] = (n-m)L_{n+m}+\frac{d}{12}n(n^2-1)\delta_{n+m,0},\\\label{Vir2}
	&&[S_n, L_m] = (n-m)S_{n+m},\\\label{Vir3}
	&&[S_n, S_m] = (n-m)L_{n+m}+\frac{d}{12}n(n^2-1)\delta_{n+m,0},
	\eea
	Here $d=2$. A natural question that arises is: what are the most general operators constructed from the modes $\alpha^i_n$ that satisfy the Virasoro algebra?
	
	\section{The generic case: The mod N operators}
	\label{ModN}
	As illustrated by the example in the previous section, the Virasoro operators naturally split into two sets: those with even indices, corresponding to the standard representation, and those with odd indices, corresponding to the new operators.
	This observation motivates the consideration of operators $S^{k}_n$ defined modulo $N$, where each operator belongs to an equivalence class labeled by $[n] = k$ for $k = 0, 1, \ldots, N-1$.
	In analogy with the $N = 2$ case discussed in the previous section, we assume that the class $S^{0}_n$ corresponds to the standard Virasoro representation
	\bea
	S_n^0=L_n=\frac{1}{2}\sum_q\alpha^i_{n-q}\alpha^i_q,~~~n=N\mathbb{Z}.
	\eea

	The most general operators in class $k$, which we denote by $V_m^k$ and $U_m^k$, can be constructed from the tensors $\epsilon^{ij}$ and $\delta^{ij}$ as follows:
	\bea
	\label{s-mostgeneric}
	&&V^k_n=\frac{1}{2}\sum_qf^k(q)\epsilon^{ij}\normord{\alpha^i_{n-q}\alpha^j_q},\\
    &&U^k_n=\frac{1}{2}\sum_qh^k(q)\normord{\alpha^i_{n-q}\alpha^i_q},
	\eea
	where $k = [n] = 1, \ldots, N-1$.
	Requiring these operators to be invariant under the transformation $q \to m - q$ imposes the following constraints:
	\bea
	\label{constraints}
	f^k(n-q)=-f^k(q),~~~h^k(n-q)=h^k(q),
	\eea
	for all $q$.
	The most general operator for $S_n^k$ is then the sum of these two operators: $S_n^k=U^k_n+V_n^k$. Let us now compute the commutators of $L_n$ with $V_n^k$ and $U_n^k$\footnote{Note that, since $k \neq 0$, normal ordering is not relevant.}.
	\bea
	&&[L_n,V^k_m]=\nn\\
    &&\frac{1}{2}\sum_q\epsilon^{ij}\Big(-(m-q)f^k(q)-f^k(q-n)(q-n)\Big)\alpha^i_{n+m-q}\alpha^j_q,\nn\\
	\eea
	To ensure the closure of the algebra under commutation, the results of the previous computation require that the following relation must hold:
	\bea
	-(m-q)f^k(q)-f^k(q-n)(q-n)=(n-m)f^k(q)
	\eea
	Since $n$ takes values in $N\mathbb{Z}$ (the index of $L_n$), it is sufficient for $f^k(q)$ to be invariant modulo $N$ to satisfy the above relation.
	The most general solution is then given by:
	\bea
	\label{fdefinition}
	f^k(q)=\sum_{w=0}^{N-1} c^k_w e^{\frac{2\pi w i}{N}q},
	\eea
	We also need to impose the constraints~\eqref{constraints}, which consequently yield the following relation:
	\bea
	\sum_{w=0}^{N-1} c^k_w e^{\frac{2\pi w i}{N}q}=-\sum_{w=0}^{N-1} c^k_w e^{\frac{2\pi w i}{N}(k-q))},
	\eea
	This condition is satisfied when the coefficients take the following form:
	\bea
	\label{cons1}
	c^k_{N-w}=-c^k_w e^{\frac{2\pi i wk}{N}},
	\eea
	It should be noted that $c^k_N=c^k_0$. The same holds for $U_n^k$, with a different set of constraints arising from~\eqref{constraints}, given as follows:
	\bea
	\label{cons2}
	d^k_{N-w}=d^k_w e^{\frac{2\pi i w k}{N}}. 
	\eea
	Therefore, the most general form of the operator $S_n^k$ that is consistent with $L_n$ is given by:
	\bea
	\label{Sdefinition}
	&&S^{k}_n=U^k_n+V^k_n\nn\\
    &&=\frac{1}{2}\sum_q\sum_{w=0}^{N-1}\Big(d^k_w e^{\frac{2\pi wi}{N}q}\alpha^i_{n-q}\alpha^i_q+ c^k_w e^{\frac{2\pi wi}{N}q} \epsilon^{ij}\alpha^i_{n-q}\alpha^j_q\Big),\nn\\
	\eea
	This is subject to the conditions~\eqref{cons1} and~\eqref{cons2}.
	As shown in Appendix~\ref{AppA}, the closedness of the Witt algebra, $[S^{k_1}_n,S^{k_1}_m]=(n-m)S^{k_1+k_2}_{n+m}$, leads to the following relations.\footnote{In the computations, we ignore normal ordering, which becomes important when $k_1 + k_2 \equiv 0 \mod N$. The central term of the Virasoro algebra can be determined consistently via the Jacobi identity.}
\begin{align}
&\sum_{\substack{w_1,w_2 \\ w_1+w_2=s\ \mathrm{mod}\ N}}
\bigl(d^{k_1}_{w_1}d^{k_2}_{w_2}-c^{k_1}_{w_1}c^{k_2}_{w_2}\bigr)
\Bigl(\frac{\zeta_{k_1}^{-w_2}+\zeta_{k_2}^{-w_1}}{2}\Bigr)=d^{k_1+k_2}_s \label{eq1}\\
&\sum_{\substack{w_1,w_2 \\ w_1+w_2=s\ \mathrm{mod}\ N}}
\bigl(d^{k_1}_{w_1}c^{k_2}_{w_2}+c^{k_1}_{w_1}d^{k_2}_{w_2}\bigr)
\Bigl(\frac{\zeta_{k_1}^{-w_2}+\zeta_{k_2}^{-w_1}}{2}\Bigr)=c^{k_1+k_2}_s \label{eq2}
\end{align}

	where $\zeta_k=e^{\frac{2\pi i k}{N}}$. Furthermore, we require that the following constraints be satisfied by $c^k_{\omega}$ and $d^k_{\omega}$:
\begin{align}
&\sum_{\substack{w_1,w_2 \\ w_1+w_2=s\ \mathrm{mod}\ N}}
\bigl(d^{k_1}_{w_1}d^{k_2}_{w_2}-c^{k_1}_{w_1}c^{k_2}_{w_2}\bigr)
\Bigl(\frac{\zeta_{k_1}^{-w_2}-\zeta_{k_2}^{-w_1}}{2}\Bigr)=0 \label{eq3}\\
&\sum_{\substack{w_1,w_2 \\ w_1+w_2=s\ \mathrm{mod}\ N}}
\bigl(d^{k_1}_{w_1}c^{k_2}_{w_2}+c^{k_1}_{w_1}d^{k_2}_{w_2}\bigr)
\Bigl(\frac{\zeta_{k_1}^{-w_2}-\zeta_{k_2}^{-w_1}}{2}\Bigr)=0, \label{eq4}
\end{align}

for all $s = 0, 1, \ldots, N-1$.
	
	Solving all of the above constraints simultaneously is not straightforward; however, in the next subsections, we present the complete set of solutions for $\mathbb{Z}_2$ and $\mathbb{Z}_3$, suggesting that solutions may exist for all $\mathbb{Z}_N$.
	\subsection{The $\mathbb{Z}_2$ case}
	\label{mod2}
	The mod 2 case provides the simplest example. Here, the class $k = 1$ is the only additional class besides the standard class $k = 0$, which corresponds to the usual representation of $L_n$ for even $n$.
	Imposing the conditions~\eqref{cons1} and~\eqref{cons2} for $N = 2$ leads to the following identities:
	\bea
	c^1_0=0,~~~d^1_1=0
	\eea 
	so that $d^1_0$ and $c^1_1$ are the only nonzero coefficients in the construction of $S_n^1$.
	All the equations~\eqref{eq1}--\eqref{eq4} are satisfied when the following relation between $d^1_0$ and $c^1_1$ holds:
	\bea
	\label{CP1}
	(d^1_0)^2+(c^1_1)^2=1.
	\eea
	
	The standard representation of $S^1_n$, which corresponds to the usual construction of the Virasoro operator $L_n$ for odd $n$, is obtained for $d^1_0 = 1$ and $c^1_1 = 0$.
	The alternative construction discussed in Section~\ref{warmup} corresponds to $d^1_0 = 0$ and $c^1_1 = 1$. Allowing the coefficients $c^1_1$ and $d^1_0$ to be complex numbers, the solutions of~\eqref{CP1} satisfy $(d^1_0)^2 + (c^1_1)^2 = 1$ defines an algebraic curve in $\mathbb{C}^2$. The space of the solutions in $\mathbb{C}^2$ is topologically equivalent to a cylinder or a sphere with two punctures. 
	
	However, let's consider the above equation as the affine equation of $X^2+Y^2=Z^2$ in the chart defined by $Z\neq0$, $d^1_0=\frac{X}{Z}$ and $c_1^1=\frac{Y}{Z}$ for some projective space with appropriate dimension. This identifies the solution space with the complex projective line $\mathbb{C}P^1$, which is topologically a sphere. This is an interesting result: there exist infinitely many alternative constructions beyond the usual Virasoro operator $L_n$ for odd $n$.
    
    \subsubsection{Equivalent solutions}
    \label{Equivalent solutions}
    It appears that we have identified the true moduli space of constructions for the $\mathbb{Z}_2$ case. However, there exist automorphisms of the current algebra
    \bea
[\alpha^i_n,\alpha^j_m]=n\delta^{ij}\delta_{n+m,0}
    \eea
which relate the two seemingly distinct solutions to each other. By introducing $\text{mod}-2$ operators $\alpha^{k i}_n$, where $n \in 2\mathbb{Z} + k$ and $k = 0,1$, the above current algebra decomposes into two commuting current algebras as follows:
\begin{equation}
\begin{aligned}
    [\alpha^{0i}_n, \alpha^{0j}_m] &= n\,\delta^{ij}\,\delta_{n+m,0}, \\
    [\alpha^{1i}_n, \alpha^{1j}_m] &= n\,\delta^{ij}\,\delta_{n+m,0}, \\
    [\alpha^{0i}_n, \alpha^{1j}_m] &= 0,
\end{aligned}
\end{equation}
Each sector admits a $\mathbb{Z}_2$ automorphism of the form $\alpha^{k i}_n \to -\alpha^{k i}_n$. Consequently, the total algebra possesses a $\mathbb{Z}_2 \times \mathbb{Z}_2$ symmetry. This observation implies that $(d^1_0,c^1_1) \sim -(d^1_0,c^1_1)$, and therefore, the space of solutions must be quotiented by an overall $\mathbb{Z}_2$ factor.

There is an interesting non-local equivalence between these algebras. The abelian Kac-Moody algebra arises from free boson theory. Suppose we have two free bosons $\tilde X^i$ and $X^i$. Then the following non-local transformation,
\begin{equation}
\label{non-local}
\tilde X(\tau,\sigma) = \frac{1 + e^{i\theta}}{2} X(\tau,\sigma) + \frac{1 - e^{i\theta}}{2} X(\tau,\sigma + \pi),
\end{equation}
where $X \equiv (X^1 + i X^2)/\sqrt{2}$, relates the above algebras to each other, with $\theta$ parametrizing the circle defined in~\eqref{CP1}. However, we do not consider such non-local transformations in this paper.
	\subsubsection{Unitarian subsector}
	\label{unitarian2}
	The above analysis demonstrates that there exists a vast number of new constructions beyond the standard Virasoro realization. However, from the conventional point of view, not all of these constructions correspond to Hermitian operators or possess a natural relation to their adjoint counterparts. To maintain consistency with the standard framework of two-dimensional conformal field theory, we assume the usual adjoint action on the $\alpha^i_n$ modes,
	\bea
	\alpha^{i\dagger}_n=\alpha^i_{-n},
	\eea
	which induces an adjoint action on the Virasoro operators $S_n$. A natural expectation is the following relation to be hold
	\bea
	\label{hermitioncondition}
	L_n^{\dagger}=L_{-n}.
	\eea
This imposes a strong constraint on the space of solutions. The even sector, represented by $S^0_n = L_n$, is by definition consistent with this adjoint action. However, the odd sector requires additional restrictions. Explicitly, one finds
	\bea
	(S^1_n)^{\dagger}=\frac{1}{2}\Big(d^{1*}_0\sum_q\alpha^{i}_{-n-q}\alpha^i_q+c^{1*}_1\sum_q\epsilon^{ij}\alpha^{i}_{-n-q}\alpha^j_q\Big),
	\eea
which is consistent with the equation~\ref{hermitioncondition}, when $d^{1*}_0=d^1_0$ and $c^{1*}_1=c^1_1$. Therefore, the space of solutions consistent with the adjoint action is  $\frac{S^1}{\mathbb{Z}_2}$, which is topologically a circle.
	\subsection{The $\mathbb{Z}_3$ case}
	Although we cannot solve Equations~\eqref{eq1}--\eqref{eq4} in closed form, they can still be solved directly without much difficulty. Beyond the $\mathbb{Z}_2$ case, it is instructive to examine the structure of the $\mathbb{Z}_3$ case and understand how it differs from $\mathbb{Z}_2$.
	
	The constraints~\eqref{cons1} and~\eqref{cons2} determine the following relations between $c^k_{\omega}$ and $d^k_{\omega}$.
	\bea
	\label{independentcoe}
	&&d^1_2=d^1_1e^{\frac{2\pi i}{3}},~~~d_2^2=d_1^2e^{\frac{4\pi i}{3}}\\
	&&c_2^1=-c^1_1e^{\frac{2\pi i}{3}},~~~c^2_2=-c_1^2e^{\frac{4\pi i}{3}},~~~c^k_0=0,
	\eea
	Working with the independent coefficients, the consistency of Equations~\eqref{eq1}--\eqref{eq4} yields the following constraints on the coefficients.
    \begin{widetext}
    \bea
&&d^1_0d^2_0+2(d_1^1d_1^2+c_1^1c_1^2)=1,\\
	&&d_1^1d_1^2-c_1^1c_1^2+e^{\frac{4\pi i}{3}}d^1_1d^2_0+e^{\frac{2\pi i}{3}}d^1_0d^2_1=0,\\
	&&e^{\frac{2\pi i}{3}} d^1_0c^2_1+e^{\frac{4\pi i}{3}}d^2_0c^1_1-(d_1^1c_1^2+c_1^1d_1^2)=0\\\label{Z3Eq3}
	&&(d^1_0)^2-e^{\frac{2\pi i}{3}}\Big((d^1_1)^2+(c^1_1)^2\Big)=d^2_0,\\\label{Z3Eq4}
	&&(d_1^1)^2-(c_1^1)^2-e^{\frac{2\pi i}{3}}d^1_0d^1_1=d_1^2,\\\label{Z3Eq5}
	&&-e^{\frac{2\pi i}{3}}d^1_0c^1_1-2d_1^1c_1^1=c_1^2,\\
	&&(d^2_0)^2-e^{\frac{4\pi i}{3}}\Big((d_1^2)^2+(c_1^2)^2\Big)=d^1_0,\\
	&&-e^{\frac{4\pi i}{3}}d^2_0d^2_1+(d_1^2)^2-(c_1^2)^2=d_1^1,\\
	&&-e^{\frac{4\pi i}{3}}d^2_0c^2_1-2d_1^2c_1^2=c_1^1,\\
	&&(d^1_0d^2_1+d^1_1d^2_0)(e^{\frac{4\pi i}{3}}-1)+(d^1_1d^2_1-c^1_1c^2_1)(e^{\frac{2\pi i}{3}}-e^{\frac{4\pi i}{3}})+d^1_1d^2_0(1-e^{\frac{2\pi i}{3}})=0,\\
	&&(d^1_1c^2_1+c^1_1d^2_1)(e^{\frac{2\pi i}{3}}-e^{\frac{4\pi i}{3}})-d^1_0c^2_1(e^{\frac{2\pi i}{3}}-1)-c^1_1d^2_0(1-e^{\frac{2\pi i}{3}})=0    
    \eea
    \end{widetext}

	The above equations appear too complicated to be solved analytically. However, one can use Equations~\eqref{Z3Eq3}--\eqref{Z3Eq5} to express $d^2_0$, $d^2_1$, and $c^2_1$ in terms of $d^1_0$, $d^1_1$, and $c^1_1$. To simplify the notation, let us define $d^1_0 \equiv x$, $d^1_1 \equiv w$, and $c^1_1 \equiv y$. Eliminating $d^2_0$, $d^2_1$, and $c^2_1$ from the equations then yields the following simplified system:
	\bea
	\label{smoothsolutions}
	&&x^3+2w^3-3e^{\frac{2\pi i}{3}}x(w^2+y^2)-6wy^2=1,\\
	&&z_i\Big(x^3+2w^3-3e^{\frac{2\pi i}{3}}x(w^2+y^2)-6wy^2=1\Big)
	\eea
	where $z_i$ is equal to $x$,$w$ or $y$.
	
	\subsubsection{Equivalent solutions}
	In the $\mathbb{Z}_3$ case, the automorphisms of the current algebra are more intricate, and one must carefully account for equivalent solutions. The current algebra remains invariant under the transformations
    \bea
    &&\alpha^{0i}_n\to-\alpha^{0i}_n,~~~\alpha^{1i}_n\to\lambda\alpha^{1i}_n,~~~\alpha^{2i}_n\to\frac{1}{\lambda}\alpha^{2i}_n,
    \eea
   for any nonzero $\lambda \in \mathbb{C}^\times$. Since the coefficients of the operators in~\eqref{Sdefinition} depend only on the residue classes modulo $N$, we only need to consider automorphisms that act identically on all members of a given class. Consequently, the automorphism group compatible with the $\mathbb{Z}_3$ grading is isomorphic to $\mathbb{C}^\times \times \mathbb{Z}_2$. Hence, two solutions $(x',w',y')$ and $(x,w,y)$ are equivalent, if there exist $\lambda\in C^{\times}$ satisfying the following relations
   \bea
   \label{equiv1}
   &&x'-\omega^2w'=\lambda(x-\omega^2w),\\\label{equiv2}
   &&x'+2\omega^2w'=\frac{1}{\lambda^2}(x+2\omega^2w),\\\label{equiv3}
   &&y'=\lambda y,
   \eea
   where $\omega=e^{\frac{2\pi i}{3}}$. Therefore, the space of inequivalent constructions is obtained by quotienting the solution space~\eqref{smoothsolutions} by the above equivalence relation. Let us define 
   \bea
   &&x_1=x-\omega^2w,\\
   &&x_2=x+2\omega^2w,\\
   &&x_3=\sqrt{3}\omega^2y,
   \eea
   and interestingly the equation~\eqref{smoothsolutions} can be written as 
   \bea
   (x_1-x_3)(x_1+x_3)x_2=1.
   \eea

   Since both $x_2$ and $\lambda$ are nonzero, we can always fix the gauge by setting $x_2 = 1$ through an appropriate choice of $\lambda$. Under this gauge fixing, the defining equation simplifies to
   \bea
   \label{gaugedfixed}
   x_1^2-x_3^2=1,
   \eea
   which admits the following parametrization:
   \bea
   x_1=\frac{t+t^{-1}}{2},~~~x_3=\frac{t-t^{-1}}{2},
   \eea
   This representation makes it clear that the space of solutions is topologically equivalent to a cylinder.
	\subsubsection{Special constructions}
    \label{specialconstructions}
	
	To obtain concrete examples of the new representations of $\mathbb{Z}_3$, we focus on the subset of solutions characterized by the vanishing of at least one of $x$, $w$, or $y$. In the zero-dimensional case, we impose two constraints on the solution space simultaneously. No solutions exist when $x = w = 0$, which implies that there are no representations for the classes $k = 1$ and $k = 2$ defined solely by the tensor $\epsilon^{ij}$ (without contributions from the $\delta^{ij}$ term). This is in contrast to the $\mathbb{Z}_2$ case, where a representation does exist for $k = 1$.
	
	Suppose that $w = y = 0$. In this case, we obtain
	\bea
	x^3=1\to d^1_0=e^{\frac{2\pi i a}{3}},
	\eea
	for $a = 0, 1, 2$. The only remaining nonzero coefficient is $d^2_0 = e^{\frac{4\pi i a}{3}}$. Substituting this solution into Equation~\eqref{Sdefinition} reproduces the standard representations of the Virasoro algebra.
	\bea
	\label{wittalgebra}
	&&[S^{k_1}_n,S^{k_2}_m]=(n-m)S^{k_1+k_2}_{n+m}+\frac{d}{12}(n^3-n)\delta^{k_1+k_2,0}\delta_{m+n,0},\nn\\
	\eea
	The Virasoro algebra~\eqref{wittalgebra} is invariant under the transformation $S^k_n \to e^{\frac{2\pi i k}{N}} S^k_n$, which is an element of the automorphism group of the Virasoro algebra. Consequently, the factors $e^{\frac{2\pi i a}{3}}$ and $e^{\frac{4\pi i a}{3}}$ can be omitted, yielding the standard representation of the Virasoro algebra.
	
	For a more nontrivial solution, let us impose the constraints $x = 0$ and $y^2 + w^2 = 0$. In this case, we obtain
	\bea
	8w^3=1\to d^1_1=\frac{1}{2}e^{\frac{2\pi i a}{3}},
	\eea
	for $a = 0, 1, 2$. The remaining nonzero independent coefficients are $d^2_1 = \frac{1}{2} e^{\frac{4\pi i a}{3}}$ and $c^2_1 = \mp \frac{1}{2} i e^{\frac{4\pi i a}{3}}$. As before, the phase factors $e^{\frac{2\pi i a}{3}}$ and $e^{\frac{4\pi i a}{3}}$ can be absorbed by rescaling the operators. Consequently, we are left with two independent solutions, corresponding to the two roots of $w^2 = -y^2$. The resulting algebras are given by
\begin{align}
S^1_n &=\frac{1}{4}\sum_q \Bigl[\bigl(e^{\frac{2\pi i}{3}q}+\omega e^{\frac{4\pi i}{3}q}\bigr)\delta^{ij} \nonumber\\
&\qquad +i\bigl(e^{\frac{2\pi i}{3}q}-\omega e^{\frac{4\pi i}{3}q}\bigr)\epsilon^{ij}\Bigr]\alpha^i_{n-q}\alpha^j_q, \label{solu1}\\
S^2_n &=\frac{1}{4}\sum_q \Bigl[\bigl(e^{\frac{2\pi i}{3}q}+\omega^2 e^{\frac{4\pi i}{3}q}\bigr)\delta^{ij} \nonumber\\
&\qquad -i\bigl(e^{\frac{2\pi i}{3}q}-\omega^2 e^{\frac{4\pi i}{3}q}\bigr)\epsilon^{ij}\Bigr]\alpha^i_{n-q}\alpha^j_q,
\end{align}
And the other solution is given by
\begin{align}
S^1_n &=\frac{1}{4}\sum_q \Bigl[\bigl(e^{\frac{2\pi i}{3}q}+\omega e^{\frac{4\pi i}{3}q}\bigr)\delta^{ij} \nonumber\\
&\qquad -i\bigl(e^{\frac{2\pi i}{3}q}-\omega e^{\frac{4\pi i}{3}q}\bigr)\epsilon^{ij}\Bigr]\alpha^i_{n-q}\alpha^j_q, \label{solu2}\\
S^2_n &=\frac{1}{4}\sum_q \Bigl[\bigl(e^{\frac{2\pi i}{3}q}+\omega^2 e^{\frac{4\pi i}{3}q}\bigr)\delta^{ij} \nonumber\\
&\qquad +i\bigl(e^{\frac{2\pi i}{3}q}-\omega^2 e^{\frac{4\pi i}{3}q}\bigr)\epsilon^{ij}\Bigr]\alpha^i_{n-q}\alpha^j_q,
\end{align}

	 Together with $S^0_n = L_n$ for $n \in 3\mathbb{Z}$, these operators define new representations of the Virasoro algebra expressed in terms of $\alpha^i_n$. The above operators are of special interest, which are compatible with the adjoint action. More precisely, for the above two solutions, we have
	
	\bea
	S^{1\dagger}_n=S^2_{-n},
	\eea
	 we analyze the space of such solutions for the $\mathbb{Z}_3$ in the subsection~\ref{unitarianZ3}. It is easy to show that there is no non-trivial solution that depends on the $\delta^{ij}$ solely. Such a space of solutions corresponds to $y = 0$:
	\bea
	\label{smoothaffine}
	x^3+2w^3-3\omega xw^2=1
	\eea
	 which admits a rational parametrization. Explicitly, a rational parametrization of the affine curve~\eqref{smoothaffine} is given by
	\bea
	x=\frac{2t^3+1}{3t^2}\omega^2,~~~w=\frac{1-t^3}{3t^2},
	\eea
	for $t\neq0$. By choosing $\lambda = \frac{1}{\omega^2 t}$ and applying the equivalence relations~\eqref{equiv1}–\eqref{equiv3}, we find that the following identities hold:
    \bea
    \lambda(x-\omega^2 w)=1,~~~\frac{1}{\lambda^2}(x+2\omega^2 w)=1,
    \eea
This implies that there are no nontrivial solutions that depend solely on $\delta^{ij}$ and the only such solution is the trivial one. In Section~\ref{Deformedalgebras}, we will demonstrate that this conclusion extends to the general $\mathbb{Z}_N$ case as well.

\subsubsection{Unitarian subsector}
	\label{unitarianZ3}
	
As in the case of the $\mathbb{Z}_2$ construction, it is of particular interest to determine the subset of constructions that remain compatible with the condition~\eqref{hermitioncondition}:
	\bea
	\label{hermitionoperators}
	S^{1\dagger}_{3n+1}=S^2_{-(3n+1)}.
	\eea 
	Since the space of constructions is determined by Eq.~\eqref{gaugedfixed}, we can explicitly express $L^1_n$ and $L^2_n$ in terms of the coordinates $x_1$ and $x_3$ (after fixing the gauge $x_2 = 1$) as follows:
\begin{align}
S^1_{3n+1} &= x_1\alpha^{i}_{3n+1-3q}\alpha^i_{3q}+\frac{1}{2}\alpha^i_{3n+1-(3q+2)}\alpha^{i}_{3q+2} \nonumber\\
&\quad -\frac{1}{2\sqrt{3}}(\omega-\omega^2)x_3\epsilon^{ij} \nonumber\\
&\qquad \times\bigl(\alpha^i_{3n+1-3q}\alpha^j_{3q}-\alpha^i_{3n+1-(3q+1)}\alpha^j_{3q+1}\bigr), \\
S^2_{3n+2} &= x_1\alpha^{i}_{3n+2-3q}\alpha^i_{3q}+\frac{1}{2}\alpha^i_{3n+2-(3q+1)}\alpha^{i}_{3q+1} \nonumber\\
&\quad -\frac{1}{2\sqrt{3}}(\omega^2-\omega)x_3\epsilon^{ij} \nonumber\\
&\qquad \times\bigl(\alpha^i_{3n+2-3q}\alpha^j_{3q}-\alpha^i_{3n+2-(3q+2)}\alpha^j_{3q+2}\bigr)
\end{align}

	where the sum over $q$ is implicit. 
    
    The unitarity condition~\eqref{hermitionoperators} requires that
    \bea
    x_1^{*}=x_1,~~~x_3^{*}=x_3,
    \eea
	which implies that the space of unitary constructions forms a hyperbola in the $(x_1, x_3)$-plane.

	\subsection{The $\mathbb{Z}_4$ case}
    Both the $\mathbb{Z}_2$ and $\mathbb{Z}_3$ cases exhibit a similar pattern. The spaces of constructions given in Eqs.~\eqref{CP1} and~\eqref{gaugedfixed} are topologically cylinders over the complex numbers. However, the unitary subsector differs between cases: for $\mathbb{Z}_2$, it corresponds to a circle, while for $\mathbb{Z}_3$, it forms a hyperbola. This observation motivates us to investigate the $\mathbb{Z}_4$ case, where new layers of complexity emerge. The number of defining equations in the $\mathbb{Z}_4$ case is considerably larger, and for clarity, we have summarized them in Appendix~\ref{AppB}. The constraints~\eqref{cons1} and~\eqref{cons2} indicate that several coefficients are not independent, leading to the following relations:
\begin{align}
d^1_3 &= i d^1_1,\; d^1_2 = 0,\; d^2_3 = -d^2_1,\; d^3_3 = -i d^3_1,\; d^3_2 = 0, \\
c^1_3 &= -i c^1_1,\; c^2_3 = c^2_1,\; c^2_2 = 0,\; c^3_3 = i c^3_1.
\end{align}

For convenience, we introduce the notations $d^1_0 \equiv x$, $d^1_1 \equiv w$, $c^1_1 \equiv y$, and an additional variable $c^1_2 \equiv u$. It is then straightforward that the space of all equations in appendix~\ref{AppB} reduces to a single equation as follows

\bea
\label{affine2}
&&-4 w^4- 4 i w^2 x^2 + x^4 + 8 w^2 y^2 - 4 i x^2 y^2 - 4 y^4 \nn\\
&&- 16 w x y u + 4 i w^2 u^2 + 2 x^2 u^2 + 4 i y^2 u^2 + u^4 = 1,\nn\\
\eea
which defines a six-dimensional hypersurface. However, as we discussed for the $\mathbb{Z}_2$ and $\mathbb{Z}_3$ cases, this hypersurface must be quotiented by the automorphisms of the current algebra that are consistent with the $\mathbb{Z}_4$ grading. The current algebra with $\mathbb{Z}_4$ grading is invariant under the following rescaling transformations:
\bea
\label{rescaling}
\alpha^{0i}_q\to -\alpha^{0i}_q,~~~\alpha^{2i}_q\to-\alpha^{2i}_q,~~~\alpha^{1i}_q\to\lambda \alpha^{1i}_q,~~~\alpha^{3i}_q\to\frac{1}{\lambda}\alpha^{3i}_q,\nn\\
\eea
where $\lambda$ is a non-zero complex parameter. Hence, the automorphism group of the current algebra consistent with the $\mathbb{Z}_4$ grading is given by $\mathbb{Z}_2\times\mathbb{Z}_2\times\mathbb{C}^{\times}$.

Analogous to the $\mathbb{Z}_3$ case, we introduce the following coordinate redefinitions:
\bea
&&x_1=x+(1+i)w,~~~x_2=x-(1+i)w,\nn\\
&&x_3=(1-i)y+u,~~~~x_4=(1-i)y-u,
\eea
With these definitions, the equation~\eqref{affine2} simplifies to the compact form
\bea
\label{affine3}
(x_1+ix_3)(x_1-ix_3)(x_2+ix_4)(x_2-ix_4)=1,
\eea
and the rescaling transformation~\eqref{rescaling} induces the following equivalence relations among the new coordinates:
\bea
x_{1}\sim\pm\lambda x_1,~~x_2\sim\frac{1}{\lambda}x_2,~~x_{3}\sim\pm\lambda x_3,~~x_4\sim\frac{1}{\lambda}x_4
\eea
Therefore, to obtain the true moduli space of constructions, the six-dimensional hypersurface must be quotiented by the group $\mathbb{Z}_2 \times \mathbb{C}^{\times}$. Since $x_2-ix_4$ cannot be zero, one can set $x_2-ix_4=1$ by an appropriate $\lambda\in \mathbb{C}^{\times}$. Then, the equation~\eqref{affine3} simplifies as follows:
\bea
\label{affine4}
(2x_2-1)(x_1+ix_3)(x_1-ix_3)=1,
\eea
which should also be quotiented by an additional $\mathbb{Z}_2$ symmetry acting as $x_1 \sim -x_1$ and $x_3 \sim -x_3$. 

%Topologically, the equation~\eqref{affine4} is equivalent to
%\bea
%(\mathbb{C}^{\times})^2\equiv T^2\times R^2,
%\eea
%which exhibits a richer topological structure compared to the $\mathbb{Z}_2$ and $\mathbb{Z}_3$ cases and can be compactified to $T^2 \times S^2$. From a geometric viewpoint, this case is also more interesting. Let us define the complex coordinates $u = x_1 + i x_3$ and $v = x_1 - i x_3$. Then, we have $x_2 = \frac{1}{2uv} \in \mathbb{C}^{\times}$, and the space satisfies the identification
%\bea
%\{ (u, v) \in \mathbb{C}^2 \mid uv \neq 0 \} \big/ (u, v) \sim (-u, -v)
%\eea
%which defines a cone (excluding the vertex point).

\subsection{The general pattern}
The analysis of the $\mathbb{Z}_2$, $\mathbb{Z}_3$ and $\mathbb{Z}_4$ cases reveals the general pattern. For a general $\mathbb{Z}_N$ grading, the corresponding current algebra  
\bea
[\alpha^{k_1 i}_n,\alpha^{k_2 j}_m]=n\,\delta^{k_1+k_2=N~mod~N}\,\delta_{n+m,0}\,\delta^{ij}
\eea
admits automorphisms that preserve the $\mathbb{Z}_N$ grading of the form  
\bea
\alpha^{k i}_n\to \lambda_k\,\alpha^{k i}_n,
\qquad 
\alpha^{(N-k)i}_n\to \lambda_k^{-1}\,\alpha^{(N-k)i}_n,
\eea
for arbitrary nonzero complex parameters $\lambda_k$. When $N$ is even, the number of complex parameters $\lambda_k$ is $\frac{N-2}{2}$ and we have an extra $\mathbb{Z}_2\times \mathbb{Z}_2$ symmetry. For odd $N$, the number of complex parameters $\lambda_k$ is $\frac{N-1}{2}$ together with an extra $\mathbb{Z}_2$ symmetry. The preceding analysis shows that the hypersurface determined by the equations~\eqref{eq1}--\eqref{eq4} should be quotiented by the symmetry group $(\mathbb{C}^{\times})^{\frac{N-2}{2}}\times \mathbb{Z}_2$ when $N$ is even, and by $(\mathbb{C}^{\times})^{\frac{N-1}{2}}$ when $N$ is odd. This demonstrates that the space of constructions for $\mathbb{Z}_{2N}$ and $\mathbb{Z}_{2N+1}$ is more or less the same over the complex numbers. However,~\eqref{CP1} and~\eqref{gaugedfixed} indicate that the unitarian subsectors, which are defined over the real numbers, may differ.

	\section{New algebras from the action principle}
	\label{equivariantsection}
	In the preceding sections, we have investigated the relationship between the $u(1)^2$ Kac--Moody algebra and the Virasoro algebra from a strictly algebraic standpoint, neglecting the question of how these algebras arise from symmetries of an underlying action. This perspective is often preferred in the physics literature. In this section, we turn to a more physically motivated problem: the construction of actions whose symmetry algebras reproduce the specific representations of the Virasoro algebra considered above.
	
	As a first step, we focus on the representation defined in Equation~\eqref{defS_n} and analyze the conditions under which it can be derived from an action principle. This example will then serve as a prototype for formulating a more general and systematic procedure for obtaining such representations. In this construction, the index $n$ associated with the operator $S_n^1$ is restricted to odd integers
	\bea
	\label{newS}
	S^1_n=\frac{1}{2}\sum_q(-1)^q\epsilon^{ij}\normord{\alpha^i_{n-q}\alpha^j_q},
	\eea  
	which couples the odd-index modes $\alpha^i_n$ to the even-index modes.
	
	To analyze the origin of this representation, let us consider a set of scalar fields $\phi^i$ defined on a flat cylinder with coordinates $(\tau,\sigma)$, where $-\infty < \tau < \infty$ and $0 \leq \sigma \leq 2\pi$. We impose the periodic boundary condition
	\bea 
	\phi^i(\tau,\sigma)=\phi^i(\tau,\sigma+2\pi),
	\eea  
	ensuring that $\phi^i$ is single-valued on the spatial circle. Under this condition, the scalar field admits the Fourier mode expansion
	\bea
	\phi^i=\sum_n a^i_n(\tau) e^{in\sigma}\equiv \phi^i_0+\phi^i_1,
	\eea
	where $\phi^i_0$ contains only the modes with $n \in 2\mathbb{Z}$, while $\phi^i_1$ contains those with $n \in 2\mathbb{Z}+1$. This decomposition naturally suggests that the operator defined in~\eqref{newS} originates from a schematic interaction terms of the form
	\bea 
	\sim\epsilon_{ij}\phi^i_{k_1}\phi^j_{k_2},
	\eea
	where $k_{1,2}=0,1$ label the two $\mathbb{Z}_2$ sectors corresponding to even and odd modes, respectively. In other words, the representation~\eqref{newS} reflects a mod-$2$ partition of the integer mode numbers.
	
	While this interpretation appears natural, it suffers from a significant drawback: the decomposition into even and odd Fourier modes relies crucially on the flat metric of the cylinder. Fourier modes only acquire an unambiguous meaning once a particular representative of the conformal class has been chosen, effectively amounting to a gauge fixing of the underlying gauge symmetries. Ideally, one would like the index $k \in \mathbb{Z}_2$ to possess an intrinsic, coordinate-independent meaning that does not depend on the specific choice of representative within the conformal class of metrics on the cylinder.
	
	In what follows, we develop a framework capable of addressing this difficulty in a systematic manner. Let $M$ be the underlying manifold and let $Diff(M)$ denote its diffeomorphism group. Given a diffeomorphism $f \in Diff(M)$,
	\bea
	f: M \to M,
	\eea
	induces a natural pullback map on the algebra of smooth complex-valued functions defined on $M$:
	\bea
	f^{*}: C^{\infty}(M,\mathbb{C}) \longrightarrow C^{\infty}(M,\mathbb{C}),
	\eea
	which acts on any $\phi \in C^{\infty}(M,\mathbb{C})$ according to
	
	\bea
	f^{*}\phi(p)=\phi(f(p)),
	\eea
	
	A further ingredient in this construction is the introduction of a group action (finite group or a finite Lie group) on $M$. More precisely, let $G$ be a group and consider a homomorphism
	\bea
	\psi: &G& \longrightarrow Diff(M), \nn\\
	&g& \longmapsto \psi_g,
	\eea
	which assigns to each $g \in G$ a diffeomorphism $\psi_g \in Diff(M)$. In this way, $G$ acts on the space of smooth functions on $M$ via the pullback induced by $\psi_g$. This viewpoint will allow us to reformulate the even/odd decomposition in terms of group representations, potentially giving the $\mathbb{Z}_2$ grading an intrinsic interpretation.
	
	We now impose the requirement that the field $\phi: M \to \mathbb{C}$ satisfy the equivariance condition
	\bea
	\label{invariantfunctionss}
	\psi^{*}_g\phi(p)=\phi\bigl(\psi_g(p)\bigr)=L(g)\phi(p),
	\eea
	where $L$ denotes a (generally linear) representation of the group $G$\footnote{For notational simplicity, we have suppressed all additional indices; the indexed case is straightforward.}. 
	
	We formalize the above property by saying that the field 
	$\phi : M \to V$ (with $V = \mathbb{C}$) is \emph{$G$–-equivariant} 
	with respect to the group action 
	\[
	\psi : G \to \mathrm{Diff}(M), \qquad g \mapsto \psi_g,
	\] 
	and a representation 
	\[
	L : G \to GL(V),
	\] 
	if the following diagram commutes for all $g \in G$:
	\[
	\begin{tikzcd}
		M \arrow[r, "\psi_g"] \arrow[d, "\phi"'] & M \arrow[d, "\phi"] \\
		V \arrow[r, "L(g)"'] & V
	\end{tikzcd}
	\]
	This expresses the condition
	$\phi(\psi_g(p)) = L(g)\,\phi(p)$ for every $p \in M$. 
	Moreover, if $f : N \to M$ is a diffeomorphism, the diagram extends as
	\[
	\begin{tikzcd}
		N \arrow[r, "\tilde{\psi}_g"] \arrow[d, "f"'] & 
		N \arrow[d, "f"] \\
		M \arrow[r, "\psi_g"] \arrow[d, "\phi"'] & 
		M \arrow[d, "\phi"] \\
		V \arrow[r, "L(g)"'] & V
	\end{tikzcd}
	\]
	where $\tilde{\psi}_g = f^{-1} \circ \psi_g \circ f$ is the conjugated action on $N$ 
	and $\tilde{\phi} = \phi \circ f$ is the pullback field. 
	The commutativity of the diagram guarantees that the equivariance property 
	is preserved under diffeomorphic changes of the base manifold.

	%This condition is consistent under diffeomorphisms between manifolds. Suppose $f$ is a diffeomorphism between two manifolds $N$ and $M$
	%\bea
	% f:N\to M,
	%\eea
	%Then one can construct a new homomorphism
	%\bea
	%\tilde\psi: &G& \longrightarrow Diff(N), \nn\\
	%&g& \longmapsto \tilde\psi_g := f^{-1} \circ \psi_g \circ f,
	%\eea
	%which defines an induced action of $G$ on $N$. In other words, the group action can be transported along the diffeomorphism $f$, ensuring that the equivariance condition~\eqref{invariantfunctionss} remains well-defined and independent of the particular representative of the diffeomorphism class of $M$.
	
	More specifically, suppose that $\phi(p)$ satisfies the equivariance condition~\eqref{invariantfunctionss}. Then its pullback under $f$, denoted by $\tilde{\phi}:= \phi \circ f$, satisfies the same relation with respect to the induced action $\tilde{\psi}_g$. Indeed, we have
	\bea
&&\tilde\psi^{*}\tilde\phi(p)=\tilde\phi(\tilde\psi_g(p))=(\phi\circ f\circ f^{-1}\circ\psi_g\circ f)(p)=\phi(\psi_g(f(p)))\nn\\
&&=\psi_g^{*}\Big(\phi\circ f(p)\Big)
	=L(g)\phi\circ f(p)=L(p)\tilde\phi(p),
	\eea
	Thus, the pullback $\tilde{\phi}$ inherits the same transformation property under the conjugated action $\tilde{\psi}_g$. Consequently, the condition~\eqref{invariantfunctionss} is invariant under diffeomorphisms between manifolds, and therefore constitutes a well-defined notion.

	Let us return to the case of the flat cylinder. The isometry group of the flat cylinder is $\mathbb{R}\rtimes O(2)$ where $\mathbb{R}$ corresponds to translations along the non-compact direction and $O(2)$ acts on the compact circle. The connected component of the identity is given by $\mathbb{R}\times S^1$. Consequently, one can always construct a homomorphism from the finite cyclic subgroup $Z_N$ into the diffeomorphism group of the flat cylinder. In fact, such a homomorphism can be defined for any manifold diffeomorphic to the flat cylinder, ensuring that the construction is robust under diffeomorphic changes of the background.

	Since $\mathbb{Z}_{N}$ is a finite Abelian group, all of its irreducible representations are one-dimensional. If $g$ denotes the generator of the group, then each irreducible representation is of the form
	\bea
	\rho_{k}: \mathbb{Z}_{N} \longrightarrow \mathbb{C}^{\times}, \qquad
	\rho_{k}(g)=e^{\frac{2\pi i k}{N}},
	\eea
	where $k=0,1,...,N-1$. We are now in a position to give a natural and intrinsic meaning to the 
	$\mathbb{Z}_N$–-index $k$ labeling the fields $\phi_k$. 
	Let $M$ be a smooth manifold with diffeomorphism group $\mathrm{Diff}(M)$, 
	and let $Z_N = \langle g \rangle$ be the finite cyclic group of order $N$. 
	Choose a group homomorphism as follows
	\[
	\psi : Z_N \longrightarrow \mathrm{Diff}(M), \qquad g \mapsto \psi_g.
	\]
	A field $\phi_k : M \to \mathbb{C}$ is said to be of \emph{sector $k$} if it is 
	$Z_N$-equivariant with respect to the character 
	\[
	\rho_k : Z_N \longrightarrow \mathbb{C}^{\times}, \qquad 
	\rho_k(g) = e^{2\pi i k / N}.
	\]
	In other words, $\phi_k$ satisfies.
	\[
	\phi_k(\psi_g(p)) = \rho_k(g)\,\phi_k(p),
	\qquad \forall p \in M,
	\]
	Which may be summarized by the commutative diagram
	\[
	\begin{tikzcd}
		M \arrow[r, "\psi_g"] \arrow[d, "\phi_k"'] & 
		M \arrow[d, "\phi_k"] \\[2pt]
		\mathbb{C} \arrow[r, "\rho_k(g)"'] & 
		\mathbb{C}
	\end{tikzcd}
	\]
	for every $g \in Z_N$. 
	
	This construction provides an intrinsic, representation–theoretic 
	interpretation of the index $k$: it labels the irreducible characters 
	of $Z_N$ under which the field transforms. 
	\subsection{Equivariant conformal field theories}
	In the previous section, we introduced the concept of $G$-equivariant functions on a manifold endowed with a group action. Here, our goal is to formulate actions constructed from such functions and to investigate whether they can give rise to conformal symmetry. Although $G$--equivariance is naturally connected to the new representations introduced in Section~\ref{ModN}, one must be cautious: these representations do not necessarily preserve conformal invariance, and their compatibility with conformal structures requires a careful analysis. As a concrete illustration, we turn to the case of $U(1)$-equivariant functions defined on a flat cylinder.
	
	\paragraph{Example: $U(1)$–-equivariant field on $\mathbb{R}\times S^1$.}
	Let $M =\mathbb{R}\times S^1$ with coordinate $\sigma \in [0,2\pi)$ for the circle. 
	Consider the group $G = U(1)$ acting by rotations
	\[
	\psi_{\alpha} : S^1 \to S^1, \qquad \psi_{\alpha}(\sigma) = \sigma + \alpha \pmod{2\pi}.
	\]
	Let $V = \mathbb{C}$ and take the representation
	\[
	L : U(1) \to \mathbb{C}^{\times}, \qquad L(e^{i\alpha}) = e^{in\alpha},
	\]
	for integer $n$. 
	Then a $U(1)$–-equivariant field $\phi : S^1 \to \mathbb{C}$ satisfies
	\[
	\phi\!\left(\sigma + \alpha\right) = e^{in\alpha}\, \phi(\sigma),
	\]
	for all $\sigma$.
	
	Now, let us consider the Polyakov action on manifolds that are topologically equivalent to the flat cylinder,
	\bea
	\label{polyakov}
	S \sim \int d^2\sigma \sqrt{-h} h^{\alpha\beta}\partial_{\alpha}\bar{\phi}\partial_{\beta}\phi ,
	\eea
	One can always construct a homomorphism from the $U(1)$ group into the diffeomorphism group of such manifolds. For definiteness, let us fix the metric to be that of the flat cylinder and take the homomorphism as in the preceding example. Suppose further that the phase space of the theory is restricted to those maps satisfying the equivariance condition
	\bea
	\phi(\tau,\sigma+\alpha) = e^{in\alpha}\phi(\tau,\sigma).
	\eea
	In this case, the only surviving mode is of the form $e^{in\sigma}$. Consequently, the full Virasoro algebra cannot be recovered from such a restricted spectrum, as the dynamics reduce essentially to a single mode.
	
	However, if we allow all possible representations, the full Virasoro algebra can in fact be recovered. This suggests that the obstruction observed in the single-mode case arises merely from an overly restrictive choice of equivariance. To address this, we turn to the study of the entire space of equivariant functions associated with a given group action. A natural question is whether a generic field can be decomposed into a sum of equivariant components, each transforming according to a different representation of the group. Such a decomposition would restore the richness of the mode expansion and, consequently, the algebraic structures associated with conformal symmetry.
	
	Suppose that $C^{\infty}(M,\mathbb{C})$ is the space of all smooth complex-valued functions on the manifold $M$. As in the previous section, we assume that there exists a homomorphism from the finite group $G$ to the diffeomorphism group of the manifold $M$ by $\psi_g: G\to Diff(M)$. We show that the space of smooth functions on the manifold $M$, namely $C^{\infty}(M,\mathbb{C})$, can be decomposed into subspaces of $\rho$-equivariant functions. For an irreducible representation $\rho$ of $G$, we define
\begin{equation}
C^{\infty}(M)_{\rho} := \bigl\{ F : M \to \mathbb{C} \text{ smooth } \big| F(\psi_{g}(x)) = \rho(g) F(x),\;\forall g \in G \bigr\},
\end{equation}

	The decomposition of $C^{\infty}(M,\mathbb{C})$ into these isotypic subspaces is obtained by means of natural projection operators. Specifically, the projection onto $C^{\infty}(M,\mathbb{C})_{\rho}$ can be expressed by averaging over the group $G$ against the character $\chi_{\rho}$ of $\rho$,
	
	\bea
	(P_{\rho} f)(x) \;:=\; \frac{d_\rho}{|G|} \sum_{g \in G} \chi_\rho(g^{-1}) \, f(\psi_{g}(x)),
	\eea
	Here $d_\rho = \dim \rho$ and $\chi_\rho = \mathrm{Tr}(\rho)$ denotes the character of $\rho$. Since $G$ is finite and abelian, every irreducible representation $\rho$ is one-dimensional, so $d_\rho = 1$ and $\chi_\rho(g) = \rho(g)$ for all $g \in G$.

	The standard orthogonality relations of characters imply the following properties:
\begin{equation}
P_\rho^{2} = P_\rho,\qquad P_\rho P_\sigma = 0\;\; \text{if } \rho \not\simeq \sigma,\qquad \sum_{\rho \in \widehat{G}} P_\rho = \mathrm{Id}_{C^{\infty}(M)}.
\end{equation}

	Here $\widehat G$ denotes the set of all inequivalent irreducible representations of $G$. Hence, the family $\{P_\rho\}_{\rho \in \widehat{G}}$ forms a complete set of mutually orthogonal projections, and one obtains the direct sum decomposition.
	\bea
	\label{decomposition}
	C^{\infty}(M,\mathbb{C}) \;\cong\; \bigoplus_{\rho \in \widehat{G}} C^{\infty}(M)_{\rho}.
	\eea

	We are now in a position to introduce the action associated with the new representation discussed in Section~\ref{ModN}. To this end, let us consider the Polyakov action~\eqref{polyakov} and expand the fields $\phi^i$ in terms of $G-equivariant$ components. More precisely, using the decomposition~\eqref{decomposition}, we write
	\bea
	\phi^i=\sum_{k=0}^{N-1}a_k\phi^i_k,
	\eea
	This expansion allows us to reformulate the Polyakov action as a sum of contributions from the various isotypic components of the field space, thereby making the role of the $G$-equivariant decomposition explicit. Then the Polyakov action takes the following form
	\bea
	\label{polyakov2}
	S \sim \int d^2\sigma \sqrt{-h} h^{\alpha\beta}\partial_{\alpha}\bar{\phi}^i_{k_1}\partial_{\beta}\phi^i_{k_2}\bar{a}_{k_1}a_{k_2},
	\eea
	where summation over the indices $k_1,k_2$ is understood. Here, the coefficients $a_{k}$ encode the expansion of the fields $\phi^i$ in terms of the $G$-equivariant components $\phi^i_k$, while the barred quantities denote complex conjugates.
	
	There is a natural generalization of the action described above, which incorporates both symmetric and antisymmetric couplings among the equivariant components of the fields. Explicitly, we propose.
\begin{equation}
\begin{aligned}
S=\frac{-1}{4\pi\alpha'} \int d^2\sigma &\sqrt{-h} h^{\alpha\beta}\partial_{\alpha}\bar{\phi}^i_{k_1}\partial_{\beta}\phi^j_{k_2} \\
&\times\Big(\delta_{ij}A_{k_1k_2}+\epsilon_{ij}B_{k_1k_2}\Big),
\end{aligned}
\label{genaction}
\end{equation}
	where $A_{k_1k_2}$ (respectively $B_{k_1k_2}$) are symmetric (respectively antisymmetric) complex coefficients. The first term, proportional to $\delta_{ij}$, is the generalized usual structure of the Polyakov action, while the second term, proportional to the antisymmetric tensor $\epsilon_{ij}$, provides a natural generalization of it. We claim that this generalized action reproduces the Virasoro algebra in a form consistent with the new algebraic structures introduced in Section~\ref{ModN}. In the next section, we illustrate this construction in detail by working out the explicit cases of the finite groups $\mathbb{Z}_2$, which already display the essential features of the general framework.
	\subsection{$\mathbb{Z}_2$ case}
	\label{actionZ2}
	In this section, we quantize the proposed action~\eqref{genaction} for the case $G=\mathbb{Z}_2$, and demonstrate that the resulting Virasoro algebra possesses the same structure in terms of the mode operators $\alpha^1_n$. For concreteness, we restrict our attention to worldsheet manifolds that are topologically equivalent to a cylinder. Since the conformal class of metrics on the cylinder is unique, we may employ the $Diff(M)\rtimes Weyl$ gauge symmetry to fix the metric to the standard flat form, as is customary in conformal gauge
	\bea
	ds^2 = d\tau^2 + d\sigma^2,
	\eea
	with $-\infty<\tau<\infty$ and $0\leq\sigma<2\pi$. Therefore, in this gauge the action~\eqref{genaction} takes the form
	\bea
	\label{genaction2}
	S=\frac{-1}{4\pi\alpha'} \int d^2\sigma \eta^{\alpha\beta}\partial_{\alpha}\bar{\phi}^i_{k_1}\partial_{\beta}\phi^j_{k_2}\Big(\delta_{ij}A_{k_1k_2}+\epsilon_{ij}B_{k_1k_2}\Big),\nn\\
	\eea
	The corresponding energy--momentum tensor is
	\bea
	&&T_{\alpha\beta}=\partial_{\alpha}\phi^i_{k_1}\partial_{\beta}\phi^j_{k_2}(\delta_{ij}A_{k_1k_2}+\epsilon_{ij}B_{k_1k_2})\nn\\
    &&-\frac{1}{2}h_{\alpha\beta}h^{\gamma\delta}\Big(\partial_{\gamma}\phi^i_{k_1}\partial_{\delta}\phi^j_{k_2}(\delta_{ij}A_{k_1k_2}+\epsilon_{ij}B_{k_1k_2})\Big)\nn\\
	\eea
	Now consider the $\mathbb{Z}_2$ symmetry. We take the homomorphism between $\mathbb{Z}_2=\{e,g\}$ and $Diff(M)$ to be
	\bea
	&&\psi :\mathbb{Z}_2\to Diff(M),\\
	&&\psi_g(\sigma)\to\sigma+\pi,~~Mod~2\pi,
	\eea
	which corresponds to a half-period shift along the spatial direction of the cylinder. The two representations of $\mathbb{Z}_2$ are $\rho(g)=1$ and $\rho(g)=-1$. Accordingly, the $\mathbb{Z}_2$--equivariant functions on the cylinder decompose into two sectors. In the trivial representation, the fields are invariant under the half-period shift, while in the non-trivial representation, they pick up a minus sign. Explicitly, they satisfy
	\bea
	&&\phi^i_0(\sigma+\pi)=\phi^i_0(\sigma),\\
	&&\phi^i_1(\sigma+\pi)=-\phi^i_1(\sigma).
	\eea 
	which have the following mode decomposition
	\bea
	\label{modeexpansion}
	\phi^i_k(\tau,\sigma)=\sum_{m=N\mathbb{Z}+k}a^i_m(\tau)e^{im\sigma}.
	\eea
	
	The action~\eqref{genaction2} can be further simplified by noting that the $\mathbb{Z}_2$-equivariant functions form orthogonal sectors with respect to the inner product defined by integration over the spatial coordinate $\sigma$. In particular, one finds
	\bea
	\int_0^{2\pi} d\sigma  \bar{\phi}^i_{k_1}(\tau,\sigma)\phi^j_{k_2}(\tau,\sigma) \sim \delta_{k_1k_2},
	\eea
	which expresses the fact that fields belonging to inequivalent representations of $\mathbb{Z}_2$ are orthogonal under the cylinder integral. As a consequence, cross-terms with $k_1\neq k_2$ vanish in the action, and the dynamics decomposes into independent $k=0$ and $k=1$ sectors.
	Therefore, as $B_{k_1k_2}$ is antisymmetric and couples different sectors, its contribution is zero. Thus, the only contribution comes from the diagonal part of $A_{k_1k_2}$. Plugging the mode expansion~\eqref{modeexpansion} into the action~\eqref{genaction2}, and using the orthogonality of the $\mathbb{Z}_2$--equivariant modes, the action reduces to
	\bea
	&&S=\frac{1}{2\alpha'}\Big(\sum_{n=even}\int d\tau(\dot{\bar a}^i_n\dot a^i_n-n^2\dot{\bar a}^i_n\dot a^i_n)A_{00}\nn\\
    &&+\sum_{n=odd}\int d\tau(\dot{\bar a}^i_n\dot a^i_n-n^2\dot{\bar a}^i_n\dot a^i_n)A_{11}\Big)
	\eea
	Thus, the theory decomposes into two independent oscillator towers: one associated with the even modes ($k=0$ sector) and the other with the odd modes ($k=1$ sector). The quantization is straightforward. The canonical momentum conjugate to $a^i_n$ is defined as
	\bea
	p^i_n = \frac{\partial L}{\partial \dot a^i_n}=\frac{A_{k_1k_2}}{2\alpha'}\dot{\bar a}^i_n,
	\eea
	where $A_{k_1k_1}$ depends on the sector we have chosen. This leads to the standard canonical commutation relations
	\bea
	\label{quantizationrules}
	[a^i_n, p^j_m] = i\delta^{ij}\delta_{n,m}.
	\eea
	
	The equations of motion for the mode coefficients follow directly from the action and take the form
	\bea
	\ddot a^i_n + n^2 a^i_n = 0,
	\eea
	for both the $k=0$ (even) and $k=1$ (odd) sectors. The general solutions are given by
	\[
	\left\{ 
	\begin{array}{l}
		a^i_n(\tau)=i\sqrt{\frac{\alpha'}{A_{00}}}\Big(\frac{\alpha^i_n}{n} e^{-in\tau}-\frac{\tilde\alpha^i_{-n}}{n}e^{in\tau}),~~n\neq0 \\
		a^i_0(\tau)=x^i_0+2\sqrt{\frac{\alpha'}{A_{00}}}\alpha^i_0\tau,~~n=0,
	\end{array}
	\right.
	\]
	for the $k=0$ sector. For the $k=1$ sector (odd modes), the same expression holds for $n\neq 0$, but by construction, there is no $n=0$ mode in this sector, since only odd Fourier modes appear. The quantization rules~\eqref{quantizationrules} induce the following commutators among the oscillator modes. For $n \neq 0$ one finds
	
	\bea
	&&[\alpha^i_n,\bar{\alpha}^j_m]=n\delta^{ij}\delta_{n,m},\\
	&&[\tilde\alpha^i_n,\bar{\tilde \alpha}^j_m]=n\delta^{ij}\delta_{n,m},
	\eea
	valid in both the $k=0$ (even) and $k=1$ (odd) sectors. For the zero mode in the trivial sector ($k=0$), the situation is different. Here, the constant mode $x^i_0$ and its conjugate momentum $\alpha^i_0$ satisfy the canonical commutation relation
	\bea
	[x^i_0,\bar{\alpha}^j_0]=i\sqrt{\frac{\alpha'}{A_{00}}}\delta^{ij},
	\eea
	which corresponds to the standard position–-momentum algebra of the degrees of freedom of the center of mass. The mode expansions of the equivariant fields $\phi^i_k(\tau,\sigma)$ then take the form

	\bea
		&&\phi^i_0=x^i_0+2\sqrt{\frac{\alpha'}{A_{00}}}\alpha^i_0\tau+i\sqrt{\frac{\alpha'}{A_{00}}}\sum_{n=2\mathbb{Z}-\{0\}}(\frac{\alpha^i_n}{n}e^{-in(\tau-\sigma)}\nn\\
        &&+\frac{\bar{\tilde a}^i_n}{n}e^{-in(\tau+\sigma)}),\\
		&&\phi^i_1=i\sqrt{\frac{\alpha'}{A_{11}}}\sum_{n=2\mathbb{Z}+1}(\frac{\alpha^i_n}{n}e^{-in(\tau-\sigma)}+\frac{\bar{\tilde a}^i_n}{n}e^{-in(\tau+\sigma)}),\nn\\
	\eea
	
	It is instructive to examine the canonical structure. If we define the conjugate momentum of $\phi^i_k(\tau,\sigma)$ by
	\bea
	P^i_k(\tau,\sigma) = \frac{A_{kk}}{4\pi\alpha'}\dot{\bar \phi}^i_k(\tau,\sigma),
	\eea
	Then the canonical commutation relations take the form 
	\bea
	[\phi^i_{k_1}(\tau,\sigma),P^j_{k_2}(\tau,\sigma)]=\frac{1}{2}i\delta^{ij}\delta_{k_1k_2}\delta(\sigma-\sigma'),
	\eea
	This shows that each $\mathbb{Z}_2$ sector contributes one-half of the usual quantization rule for a free boson. In other words, the total phase space of the theory naturally decomposes into two independent halves, corresponding to the $k=0$ (periodic) and $k=1$ (antiperiodic) sectors.
	
	Let us now return to the Virasoro algebra. By definition, the Virasoro generators $L_m$ are obtained as the Fourier coefficients of the energy–momentum tensor. Working in light–cone coordinates $\sigma^{\pm} = \tau \pm \sigma$, the right–-moving component of the stress tensor takes the form
	\bea
	T_{--}&=&\partial_{-}\bar{\phi}^i_{k_1}\partial_{-}\phi^j_{k_2}(\delta_{ij}A_{k_1k_2}+\epsilon_{ij}B_{k_1k_2})\nn\\
    &\equiv& 2\sqrt{\frac{\alpha^{'2}}{A_{00}A_{11}}}\sum_{m\in\mathbb{Z}}L_me^{-2im\sigma^-}
	\eea
	If we define the complex conjugate of the oscillator modes by $\bar{\alpha}^i_n=\alpha^i_{-n}$, then the Virasoro generators can be written in terms of the oscillator bilinears. Explicitly, they take the form
	\bea
	L_{2n}&=&\frac{1}{2}\Big(A_{00}\sum_q \alpha^i_{2n-2q}\alpha^i_{2q}+A_{11}\sum_q \alpha^i_{2n-(2q+1)}\alpha^i_{2q+1}\Big),\nn\\
	L_{2n+1}&=&\frac{1}{2}\Big(A_{01}\sum_q\alpha^i_{2n+1-q}\alpha^i_q+B_{01}\sum_q (-1)^q\alpha^i_q\alpha^j_{2n+1-q}\epsilon_{ij}\Big).\nn\\
	\eea
	
	Focusing on the Witt (centreless Virasoro) part of the commutator $[L_{2n}, L_{2m}]$, closure of the algebra imposes algebraic constraints on the coupling coefficients
	\bea
	A^2_{00}=A_{00},~~~A^2_{11}=A_{11},
	\eea
	the standard solution in section~\ref{mod2} is correspond to $A_{00}=A_{11}=1$ with the following constraints
	\bea
	A^2_{01}+B^2_{01}=1,
	\eea
	which comes from the closedness of the odd part of the algebra.
	
	It is interesting that, by using the $G$-equivariance techniques, we uncover a richer structure than the purely algebraic constructions of section~\ref{ModN}. Not only does the usual representation of $L_{2n}$ (used as our starting point in section~\ref{ModN}) remain a valid representation, but one can also express $L_{2n}$ entirely in terms of the even or the odd oscillator towers. Explicitly,
	\bea
	\label{newcons}
	&&L_{2n} \;=\; \tfrac{1}{2} \sum_{q \in \mathbb{Z}} 
	\alpha^i_{\,2n-2q}\,\alpha^i_{\,2q}
	\quad \text{or} \quad
	\tfrac{1}{2} \sum_{q \in \mathbb{Z}} 
	\alpha^i_{\,2n-(2q+1)}\,\alpha^i_{\,2q+1}.\nn\\
\eea
	
	The above analysis reveals that the underlying mathematical structure governing these constructions is $G$-equivariance. This concept can be naturally generalized to more intricate settings, such as non-abelian Kac-Moody algebras, thereby offering alternative realizations of the Virasoro algebra beyond the conventional Sugawara construction~\cite{PhysRev.170.1659}.
	\section{Deformations of the Virasoro-Kac-Moody like algebras}
	\label{Deformedalgebras}
	We analyzed the degree of flexibility in the relation between the abelian $U(1)^D$ Kac-Moody algebra and the Virasoro algebra. This analysis reveals additional possible realizations beyond the standard Sugawara construction for $D=2$, thereby suggesting a broader class of Virasoro–Kac–Moody–type algebras. As a concrete illustration, in the $\mathbb{Z}_2$ case we have the following Lie algebra,\footnote{For coherence with the notation used in the literature, we denote $j^i_n \equiv \alpha^i_n$.}
	\bea
    \label{KacMoodyvir}
	&&[j^i_n,j^j_n]=n\delta^{ij}\delta_{n+m},\\
	&&[S^1_n,j^i_m]=-m aj^i_{m+n}+b\epsilon^{ij}m(-1)^mj^j_{m+n},\\
	&&[L_n,j^i_m]=-m j^i_{m+n},\\
	&&[S^{k_1}_n,S^{k_1}_m]=(n-m)S^{k_1+k_2}_{n+m}+\frac{c}{12}\delta^{k_1+k_2,0}n(n^2-1)\delta_{n+m,0},\nn\\
	\eea
	where the condition $k_1 + k_2 = 0$ is understood modulo $N$ and as usual $L_n=S^0_n$.\footnote{It should be emphasized that the indices $k$ labeling $S^{k}_n$ correspond to the $\mathbb{Z}_N$–grading and must not be confused with the indices $i,j$ associated with the $U(1)^2$ Kac-Moody algebra.} The parameters $a$ and $b$ satisfy the algebraic relation~\eqref{CP1}. To the best of the author's knowledge, such a two-parameter generalization of the standard $U(1)^2$ algebra has not been mentioned in the literature.

\bigskip

\noindent
One can show that the algebra introduced above cannot be transformed into the 
standard Virasoro--Kac--Moody algebra by any redefinition of the generators. 
This observation motivates the need for a precise notion of when two such algebras 
should be regarded as equivalent. 
In order to formalize this, let us define the exact meaning of an \emph{isomorphism} 
between such structured algebras.

\bigskip

\noindent
\textbf{Definition.} \\
Let $V$ be a vector space over a field $\mathbb{C}$, equipped with a basis 
$\{T_n\}_{n\in \mathbb{Z}}$. 
Although this basis is not unique for the vector space, we require that it has a special property. We assume that the vector space carries additional 
structure in the form of an equivalence relation on the index set $\mathbb{Z}$ defined by
\[
n_1 \sim n_2 \quad \Longleftrightarrow \quad n_1 \equiv n_2 \pmod{N},
\]
for some fixed positive integer $N$. 
The equivalence relation on the index set $\mathbb{Z}$ partitions the basis $\{T_n\}_{n\in\mathbb{Z}}$ into $N$ disjoint families. Each family spans a subspace $V_k$, and the vector space decomposes as a direct sum
\[
V = \bigoplus_{k=0}^{N-1} V_k
\]
where 
\[
V_k = \operatorname{span}\{T_{k+mN} \mid m \in \mathbb{Z}\}.
\]

Similarly, for another vector space $W$ with basis $\{S_n\}_{n\in\mathbb{Z}}$, we write
\[
S_{n_1}\sim S_{n_2} \quad \text{if and only if} \quad n_1\sim n_2 \pmod{N}.
\]

Let $W$ be another Lie algebra with basis $\{S_n\}_{n\in\mathbb{Z}}$ and an equivalence 
relation $\sim'$ on its index set. 
A pair $(\phi,\sigma)$ is said to define an 
\emph{equivariant Lie algebra isomorphism}
\[
(\phi,\sigma): (V,\{T_n\},\sim) \longrightarrow (W,\{S_n\},\sim')
\]
if the following conditions hold:
\begin{enumerate}
\item $\phi:V\to W$ is a Lie algebra homomorphism, i.e.
\[
\phi([T_n,T_m])=[\phi(T_n),\phi(T_m)] \qquad \forall n,m\in\mathbb{Z}.
\]
\item $\sigma:\mathbb{Z}\to\mathbb{Z}$ is a group homomorphism, typically of the form 
$\sigma(n)=a n$ for some integer $a$, encoding the relabeling of the basis indices.
\item The map $\phi$ acts on the basis by
\[
\phi(T_n)=S_{\sigma(n)}.
\]
\item The equivalence relations are compatible in the sense that
\[
n_1 \sim n_2 \pmod{N} \quad \Longrightarrow \quad \sigma(n_1) \sim' \sigma(n_2).
\]
\end{enumerate}

If, in addition, $\phi$ is bijective and $\sigma$ is invertible 
(as a group automorphism of $\mathbb{Z}$), then $(\phi,\sigma)$ defines an isomorphism of 
structured Lie algebras,
\[
(V,\{T_n\},\sim) \;\cong\; (W,\{S_n\},\sim').
\]

Such a definition imposes a strong constraint on the homomorphism $\sigma$. Suppose that the above algebra is isomorphic to the standard $U(1)^2$ Kac--Moody algebra, generated by $(\tilde j^i_n, \tilde S^k_n)$. Then, there exists a pair $(\phi, \sigma)$ characterizing this isomorphism, defined as follows:
\bea
\label{algebraisomorphism}
\phi([j^i_n,j^j_m])&=&[\phi(j^i_n),\phi(j^i_m)]=[\tilde j^i_{\sigma(n)},\tilde j^j_{\sigma(m)}]\nn\\
&=&\sigma(n)\delta^{ij}\delta_{\sigma(n)+\sigma(m),0}\delta^{\tilde k_1+\tilde k_2,0},\nn\\
\eea
where $\tilde k_1$ ($\tilde k_2$) denote the equivalence classes of $\sigma(n)$ and $\sigma(m)$, respectively. The condition~\eqref{algebraisomorphism} then leads to the following relation:
\bea
\sigma(n)=n.
\eea
Therefore, the $\mathbb{Z}_N$ grading is invariant under the map $\sigma$. The most general admissible transformation compatible with the $\mathbb{Z}_N$ grading takes the form~\footnote{It is important to emphasize that, because we have restricted our attention to local field transformations (see~\ref{Equivalent solutions}), we apply the same rotation tensor $A^{ij}$ to all $Z_N$ classes.}
\bea
\tilde j^i_n=\lambda_{[n]}A^{ij}j^j_{n},~~~~~~A^{ik}A^{jl}\delta^{kl}=\delta^{ij}
\eea
where $A^{ij}$ is an orthonormal matrix and $\lambda_{[n]}$ is an automorphism of the current algebra that depends only on the equivalence class $[n]$.

For the $\mathbb{Z}_2$ case, we focus on the odd sector, where $m \in 2\mathbb{Z} + 1$, 
\bea
\label{virasorokascMoody1}
&&[\tilde{j}^i_n, \tilde{j}^j_m] = n \delta^{ij} \delta_{n+m},\\
\label{virasorokascMoody2}
&&[S^k_n, \tilde{j}^i_m] = -m \tilde{j}^i_{m+n}.
\eea
The first relation is just $A A^{T} = 1$, while the second leads to $(a-1)A = bA\epsilon$. Taking the determinant of the latter relation gives
\bea
(a - 1)^2 = b^2,
\eea
which is consistent with $a^2 + b^2 = 1$ only for $a = 1$ and $b = 0$. Therefore, for any nontrivial choice of $(a, b)$, the resulting algebra is distinct from the standard Virasoro-Kac-Moody algebra~\eqref{virasorokascMoody1}-\eqref{virasorokascMoody2}.

It is natural to ask whether there exist non-trivial constructions beyond the trivial construction that depend solely on $\delta^{ij}$. In the $\mathbb{Z}_3$ case, we showed in section~\ref{specialconstructions} that no such additional constructions exist. Here, we provide a sketch of the argument indicating that the same conclusion holds for general $\mathbb{Z}_N$. Let us define
\bea
j^{k\bf{i}}_n\equiv j^{\bf{i}}_n, \quad \text{$[n]=k=0,1,...,N-1$~Mod $\mathbb{Z}_N$}
\eea
Then, the $S^1_{Nn+1}$ can be written as 
\begin{align}
\label{deformation}
S^1_{Nn+1} &= \sum_{k=0}^{P} a_{k} j^{(N+1-k)\mathbf{i}}_{Nn+1-(Nq+k)} j^{(k)\mathbf{i}}_{Nq+k} \nonumber\\
&\equiv \Bigl( x_1 j^{0}_{Nn+1-(nq+1)}j^{1}_{Nq+1} \nonumber\\
&\qquad + x_2 j^{N-1}_{Nn+1-(Nq+2)}j^{2}_{Nq+2} + \cdots \Bigr),
\end{align}
where the sum over $q$ is implicit, and the relation between $x_k$ and $a_k$ can be easily derived. In the above expansion, we have $P=\frac{N-2}{2}$ when $N$ is even, and $P=\frac{N-1}{2}$ when $N$ is odd. Using the automorphism group of the current algebra, we obtain the following equivalence relation for the variables $x_k$ when $N$ is even:
\bea
\label{equivalence}
x_a\sim\frac{\epsilon_a\lambda_{a}}{\lambda_{a-1}}x_a,~~a=1,...,\frac{N}{2},~~~~\lambda_0=\lambda_{\frac{N}{2}}=1,
\eea
where $\epsilon_a = \pm 1$ for $a=1$ and $a=\frac{N}{2}$, and is equal to $1$ otherwise.  
The only invariant combination of the above coefficients is $x_1x_2...x_{\frac{N}{2}}$. Therefore, the hypersurface resulting from the equations~\eqref{eq1}--\eqref{eq4} is given by
\bea
(x_1x_2...x_{\frac{N}{2}})^2=1.
\eea
Since the equivalence relations~\eqref{equivalence} introduce 
$\frac{N}{2}-1$ variables $\lambda_1,\ldots,\lambda_{\frac{N}{2}-1}$, 
we can set 
\[
x_2 = x_3 = \cdots = x_{\frac{N}{2}} = 1,
\]
which reduces the hypersurface equation to
\[
x_1^2= 1.
\]
which leads to $x_1 = \pm 1$. However, using the $\mathbb{Z}_2$ symmetry encoded in 
$\epsilon_a$, the minus solution can be mapped to the trivial one. A similar argument 
applies for odd $N$, with minor modifications.
\subsection{Operator product expansion of deformed algebras}
It is instructive to examine the operator product expansion (OPE) between the energy–momentum tensor and the currents. The space of all possible deformed algebras is vast; however, for algebras that depend solely on $\delta^{ij}$, it is straightforward to determine their general structure. As is standard in the literature, we consider the OPE relations on the complex plane rather than on the cylinder, using the conformal map
\bea
z = e^{\tau + i\sigma}.
\eea
Under this transformation, the $\mathbb{Z}_N$ action on the cylinder, $\sigma \to \sigma + \frac{2\pi}{N}$, becomes
\bea
z \to e^{\frac{2\pi i}{N}} z.
\eea

The $\mathbb{Z}_N$–equivariant functions $X_k(\tau,\sigma)$ and their derivatives with respect to $\sigma$ exhibit the same equivariance property on the cylinder. However, on the complex plane, the holomorphic $\mathbb{Z}_N$–equivariant function
\bea
X_k(e^{\frac{2\pi i}{N}} z) = e^{-\frac{2\pi i k}{N}} X_k(z)
\eea
induces a different equivariance behavior on its holomorphic derivative. More precisely,
\bea
&&J^k(e^{\frac{2\pi i}{N}} z) = e^{-\frac{2\pi i}{N}} e^{-\frac{2\pi i k}{N}} J^k(z),
\qquad J^k = \partial_z X_k(z),\nn\\
\eea
which can be generalized to a holomorphic primary field $\phi$ of conformal weight $h$ as follows
\bea
\phi^k(e^{\frac{2\pi i}{N}} z) = e^{-\frac{2\pi ih}{N}} e^{-\frac{2\pi i k}{N}} \phi^k(z).
\eea

In quantum field theory, conserved charges act as the generators of the symmetry algebra and determine how operators such as $\phi$ transform,
\bea
\delta \phi^k = [Q, \phi^k].
\eea
 Specifically, in conformal field theories, the holomorphic part of the charge is expressed as
\bea
Q = \frac{1}{2\pi i} \sum_{k_1+k_2+1=N}\oint_C dz  T^{k_1}(z) \epsilon^{k_2}(z),
\eea
where we have used the $\mathbb{Z}_N$-equivariance decomposition of the energy-momentum tensor and functions into equivariant parts. The sum is understood in way that be compatible with the $\mathbb{Z}_N$ invariance of charge $Q$.

It is straightforward to verify that the deformed algebra~\eqref{deformation} gives rise to the following operator product expansions (OPEs) between the currents and the energy–momentum tensor\footnote{We denote the $U(1)^2$ index by $\boldsymbol{i}$ in order to distinguish it from 
the $\mathbb{Z}_N$ grading indices $k$.
}:
\begin{widetext}
\bea
&&J^{k_1\bold{i}}(z)J^{k_2\bold{j}}(w)=\frac{\delta^{k_1+k_2=0}}{(z-w)^2}\delta^{\bold{ij}},\\\label{current-energy}
&&T^{k_1}(z)J^{k_2\bold{i}}(w)=\Big(a\delta^{\bold{ij}}+b(-1)^{k_2}\epsilon^{\bold{ij}}\Big)\Big(\frac{J^{(k_1+k_2)\bold{j}}(\omega)}{(z-w)^2}+\frac{\partial_{\omega}J^{(k_1+k_2)\bold{j}}(\omega)}{z-w}\Big),\\
&&T^{k_1}(z)T^{k_2}(w)=\frac{c\delta^{k_1+k_2=0}}{2(z-w)^4}+\frac{2T^{k_1+k_2}(\omega)}{(z-w)^2}+\frac{\partial_{\omega}T^{k_1+k_2}(\omega)}{z-w},
\eea    
\end{widetext}

where $J^{k\bold{i}}(z)=\sum_n z^{-n-1}j^{k\bold{i}}_n$ and $T^{k}(z)=\sum_n z^{-n-2}S^k_n$.

As we observe, the standard OPE between the currents and the energy–momentum tensor is deformed by the matrix $(a\delta^{\bold{ij}}+b(-1)^{k_2}\epsilon^{\bold{ij}})$ appearing in~\eqref{current-energy}. Moreover, the presence of this deformation modifies the usual notion of primary fields. Consequently, one can define primary fields within the deformed $\mathbb{Z}_N$-equivariant algebra as follows:

\paragraph{Definition.}
A field $\phi^{\bold{i}}(z)$ with the $\mathbb{Z}_N$ decomposition $\{\phi^{k\bold{i}}(z)\}$ is called \emph{primary} with conformal dimensions $h$ if the operator product expansions of the decompositions of energy--momentum tensor $\{T^{k}(z)\}$ with $\phi^{k\bold{i}}(z)$ take the form
\bea
&&T^{k_1}(z)\phi^{k_2\bold{i}}(w)=\nn\\
&&\Big(a\delta^{\bold{ij}}+b(-1)^{k_2}\epsilon^{\bold{ij}}\Big)\Big(
\frac{h\,\phi^{(k_1+k_2)\bold{j}}(w)}{(z-w)^2} \;+\; \frac{\partial_w\phi^{(k_1+k_2)\bold{j}}(w)}{z-w}\Big).\nn\\
\eea

It is interesting how the parameters $(a,b)$ affect the representations of the above Kac-Moody-Virasoro algebra. 

\section{Concluding remarks}
We may be interested in replacing the epsilon tensor~$\epsilon^{ij}$ with a generic antisymmetric background tensor $B^{ij}$ and extending the construction to arbitrary dimensions. Aside from issues related to Lorentz invariance, it is straightforward to show that the general case does not differ significantly from the results obtained above. For a real non-singular even dimensional $2n\times2n$ antisymmetric matrix $B^{ij}$, there is a real orthogonal $2n\times2n$ matrix $U$ such that~\cite{Zumino:1962smg,becker,greub2012linear}
	\bea
	&&U^{T}BU\nn\\
    &&=\text{diag}\Bigg(
	\begin{bmatrix}
		0  &  m_1      \\
		-m_1  &  0     
	\end{bmatrix}
	,
	\begin{bmatrix}
		0 &  m_2      \\
		-m_2  &  0      
	\end{bmatrix} 
	,\dots,
	\begin{bmatrix}
		0 &  m_n      \\
		-m_n  &  0      
	\end{bmatrix} 
	\Bigg)\nn\\
	\eea
	So the $2n\times2n$ even-dimensional case reduces to a bunch of $2\times2$ matrices which can be studied in a similar way to the previous sections. The antisymmetric matrices are singular in odd dimensions, and the following theorem is helpful for odd-dimensional antisymmetric and singular even-dimensional matrices :
	
\textbf{Theorem:}
	If $B$ is a real singular anti-symmetric $d\times d$ matrix of rank $2n$ where $d$ is even or odd and $d>2n$, then there exist a real orthogonal $d\times d$ matrix $U$ such that 
\begin{equation}
\begin{aligned}
U^{T}BU = \text{diag}\Bigg( &
\begin{pmatrix}
0 & m_1 \\
-m_1 & 0
\end{pmatrix},
\begin{pmatrix}
0 & m_2 \\
-m_2 & 0
\end{pmatrix}, \\
& \ldots,
\begin{pmatrix}
0 & m_n \\
-m_n & 0
\end{pmatrix},
\mathcal{O}_{d-2n}
\Bigg),
\end{aligned}
\end{equation}
where $\mathcal{O}_{d-2n}$ is a $(d-2n)\times(d-2n)$ block of zeros. Therefore, the general case is similar to the previous sections, and singular matrices have some zeros which can be treated in a similar manner as the conventional case.
\section{Conclusion}

In this paper, we have introduced and analyzed a new family of Virasoro–Kac–Moody algebras constructed from $Z_N$ gradings of the mode indices of the Virasoro generators. We started with a concrete construction in the $Z_2$ case, using the antisymmetric tensor $\epsilon^{ij}$. This extra invariant tensor exists in $D=2$ to build sets of Virasoro generators that differ for even and odd Fourier modes $n$. Afterwards, this construction was extended naturally to the general case of $Z_N$ gradings.

Partitioning the Virasoro generators into $N$ distinct classes based on the value of their mode index $n$ modulo $N$, and requiring that these new operators form a closed Virasoro algebra, imposes a set of algebraic constraints on their defining coefficients. Solving these constraints revealed the space of all possible constructions. We then solved these constraints for the special cases $N=2$, $N=3$, and $N=4$ to reveal the structure of the corresponding solutions. For the $N=2$ case, by separating the Virasoro operators into even and odd classes, we showed that the construction for the odd-indexed operators is not unique; in fact, there is an infinite family of valid constructions. The space of these solutions is described by the simple algebraic relation $x^2+y^2=1$, which, in $\mathbb{C}^2$, is topologically equivalent to a cylinder. It is shown that $N=3$ has a similar structure such as $N=2$. For the $N=4$ case, the space of all possible Virasoro generators corresponds to a four-dimensional manifold, $T^2\times S^2$. We have argued that the general situation for $\mathbb{Z}_{2N}$ and $\mathbb{Z}_{2N+1}$ is similar and shares the essential features, as both have the same dimensionality.

Shifting our perspective from a purely algebraic view toward a more physical one, we introduced Lagrangians with underlying symmetries that correspond to the aforementioned Virasoro–Kac–Moody algebras. Using the concept of $\mathbb{Z}_N$-equivariance, we clarified the meaning of $Z_N$ mode separation on a general manifold. The fields in these theories were classified based on how they transform under the action of a finite symmetry group $\mathbb{Z}_N$ on the underlying worldsheet manifold. This classification was thereby tied to a geometric symmetry of the space itself, making it an intrinsic, coordinate-independent property. Following this logic, we explicitly constructed the $Z_2$ case by quantizing the theory described by such an action and derived the new Virasoro generators. The final expressions were precisely the same as those found through the purely algebraic methods. This demonstrates that the $\mathbb{Z}_N$-equivariant action principle is indeed the correct physical foundation for these new algebras.

By modifying the Virasoro generators, the relation between the Kac–Moody currents and the Virasoro generators changed fundamentally. This led to the creation of entirely new, deformed versions of the Virasoro–Kac–Moody algebra. For the $Z_2$ case, we explicitly wrote down a new Lie algebra that depends on two parameters $a$ and $b$, which satisfy an algebraic relation. We also showed that such novel algebras cannot be transformed into the usual ones by any redefinition of the generators.
\section*{Acknowledgments}
The authors thank Hessamaddin Arfaei and Shahin Rouhani for their valuable comments.

	\appendix
	\section{Algebraic preliminaries: The $\mathbb{Z}_N$ commutators}
	\label{AppA}
	In this section, we derive the basic algebraic equations~\eqref{eq1}--\eqref{eq4}, which form the foundation for all computations in the first part of the paper.  
	
	Integer numbers can be partitioned into $N$ classes when considered modulo $N$. Accordingly, the Virasoro generators can be grouped into these $N$ classes, which we denote by $S^k_n$, where the index $k = 0,1,\dots,N-1$ specifies the class. We choose the trivial class $k=0$ to correspond to the usual Virasoro operators:
	\bea
	S^0_n=L_n=\frac{1}{2}\sum_q\alpha^i_{n-q}\alpha^i_q,\qquad n = N\mathbb{Z}, ,
	\eea
	We now try to fix the most generic form of other $S^k_n$ for $k=1,...,N-1$. Let us consider the action of these modes on the following family of generators:  
	\bea
	\label{s-mostgeneric1}
	V^k_m = \frac{1}{2}\sum_q f^k(q)\, \epsilon^{ij}\, \alpha^i_{m-q}\alpha^j_q.
	\eea
	For simplicity, we ignore normal ordering subtleties at this stage, which can be determined by the Jacobi identity. Therefore, we obtain
	
	\bea
	&&[L_n,V^k_m]=\frac{1}{2}\sum_qf^k(q)\epsilon^{ij}[L_n,\alpha^i_{m-q}\alpha^j_q]\nn\\
    &&=\frac{1}{2}\sum_qf(q)\epsilon^{ij}\Big([L_n,\alpha^i_{m-q}]\alpha^j_q+\alpha^i_{m-q}[L_n,\alpha^j_q]\Big)\nn\\
	&&=\frac{1}{2}\sum_qf^k(q)\epsilon^{ij}\Big(-(m-q)\alpha^i_{n+m-q}\alpha^j_q-q\alpha^i_{m-q}\alpha^j_{n+q}\Big)\nn\\
	&&=\frac{1}{2}\sum_q\epsilon^{ij}\Big(-(m-q)f^k(q)-f^k(q-n)(q-n)\Big)\alpha^i_{n+m-q}\alpha^j_q\nn\\
	\eea
	where the equivalence class of $m$ modulo $N$ is denoted by $[m]=k$.  
	For the closure of the algebra, these generators must satisfy the consistency condition
	\bea
	&&-(m-q)f^k(q)-f^k(q-n)(q-n)=(n-m)f^k(q),\nn\\
	\eea
	As $n$ is restricted to values in $N\mathbb{Z}$, it is sufficient that the functions $f^k(q)$ be invariant modulo $N$ to satisfy the above closure condition.  
	Therefore, the most general solution can be expressed in the form
	\bea
	\label{fdefinition1}
	f^k(q)=\sum_{w=0}^{N-1} c^k_w e^{\frac{2\pi w i}{N}q}
	\eea
	
	Since $f^{k}$ is a $\mathrm{mod}\,N$ invariant function, the index $w$ runs over the range $w=0,\dots, N-1$. An additional constraint arises from the transformation $m-q=u$ applied to~Equation~\eqref{s-mostgeneric1}, which leads to the following relation
	
	\bea
	f^k(q)=-f^k(m-q),
	\eea
	As $m$ in Equation~\eqref{s-mostgeneric1} takes values of the form $m \in N\mathbb{Z}+k$, with $k=1,\dots, N-1$, and since $f^k(q)$ is $\mathrm{mod}\,N$ invariant, the constraint simplifies to
	\bea
	f^k(q)=-f^k(k-q),
	\eea
	which leads to the following relation 
	\bea
	\sum_{w=0}^{N-1} c^k_w e^{\frac{2\pi w i}{N}q}=-\sum_{w=0}^{N-1} c^k_w e^{\frac{2\pi w i}{N}(k-q))}
	\eea
	This condition is satisfied provided the coefficients obey the following relation:
	\bea
	\label{cons11}
	c^k_{N-w}=-c^k_w e^{\frac{2\pi i wk}{N}},
	\eea
	where we assume that $c^k_N=c^k_0$. This provides the most general form of the coefficients ensuring the closedness condition $[L_n, S^k_m]=(m-n)S^k_{m+n}$. Let us focus on  another class of operators as 
	\bea
	\label{newhoperators1}
	U^k_n=\frac{1}{2}\sum_qh^k(q)\alpha^i_{m-q}\alpha^i_q,
	\eea
	By the same reasoning as before, the function $h^k(q)$ can be expanded in the form 
	\bea
	\label{hdefinition1}
	h^k(q)=\sum_{w=0}^{N-1}d^k_we^{\frac{2\pi wi}{N}q},
	\eea
	where the coefficients $d^k_w$ satisfy the constraint
	\bea
	\label{cons21}
	d^k_{N-w}=d^k_w e^{\frac{2\pi i w k}{N}},
	\eea
	and the identification $h^k_N = h^k_0$ is assumed, ensuring periodicity modulo $N$. Therefore, the most general form of an operator consistent with the action of $L_n$ is given by
	\bea
	\label{Sdefinition1}
	&&S^{k}_n=U^k_n+V^k_n\nn\\
    &&=\frac{1}{2}\sum_q\sum_w\Big(d^k_w e^{\frac{2\pi wi}{N}q}\alpha^i_{n-q}\alpha^i_q+ c^k_w e^{\frac{2\pi wi}{N}q} \epsilon^{ij}\alpha^i_{n-q}\alpha^j_q\Big),\nn\\
	\eea
	together with the constraints~\eqref{cons11} and~\eqref{cons21} on the coefficients $c^k_w$ and $d^k_w$. Our goal is to determine the constants $c^k_w$ and $d^k_w$ in such a way that the resulting operators $S^k_n$ satisfy the desired algebra
	\bea
	\label{Zngeneralization1}
	[S^{k_1}_n,S^{k_2}_m]=(n-m)S^{k_1+k_2}_{n+m}.
	\eea
	Using the form~\eqref{s-mostgeneric1}, the commutator can be written as
	\bea
	&&[V^k_n,\alpha^i_m]=-\frac{1}{2}m\Big(f^k(n+m)-f^k(-m)\Big)\epsilon^{ij}\alpha^j_{n+m},\nn\\
	\eea
	and the commutator of $[V^{k_1}_n,V^{k_2}_m]$ is then becomes
	\bea
	&&[V^{k_1}_n,V^{k_2}_m]=\nn\\
    &&\frac{1}{2}\sum_q\epsilon^{ij}f^{k_2}(q)\left([V^{k_1}_n,\alpha^i_{m-q}]\alpha^j_q+\alpha^i_{m-q}[V^{k_1}_n,\alpha^j_q]\right)\nn\\
	&&=\frac{1}{2}\epsilon^{ij}f^{k_2}(q)\Bigg(-\frac{1}{2}(m-q)\Big(f^{k_1}(n+m-q)\nn\\
    &&-f^{k_1}(q-m)\Big)\epsilon^{ik}\alpha^k_{n+m-q}\alpha^j_q\nn\\
	&&-\frac{1}{2}q\alpha^i_{m-q}\Big(f^{k_1}(n+q)-f^{k_1}(-q)\Big)\epsilon^{jk}\alpha^k_{n+q}\Bigg)
	\eea
	By redefining $n+q\to q$ in the second term, we finally have
	
	\bea
	\label{vcommutator1}
	&&[V^{k_1}_n,V^{k_2}_m]\nn\\
    &&=\Biggl(-\frac{1}{4}(m-q)f^{k_2}(q)\Big(f^{k_1}(n+m-q)-f^{k_1}(q-m)\Big)\nn\\
	&&+\frac{1}{4}f^{k_2}(q-n)(q-n)\Big(f^{k_1}(q)-f^{k_1}(n-q)\Big)\Biggl)\alpha^i_{n+m-q}\alpha^i_q,\nn\\
	\eea
	Moreover, the commutator of $[U^{k_1}_n,U^{k_2}_m]$ is given by
	
	\bea
	\label{ucommutator1}
	&&[U^{k_1}_n,U^{k_2}_m]\nn\\
    &&=-\frac{1}{4}\Biggl((m-q)h^{k_2}(q)\Big(h^{k_1}(q-m)+h^{k_1}(n+m-q)\Big)\nn\\
	&&+(q-n)h^k(q-n)\Big(h^{k_1}(q)+h^{k_1}(n-q)\Big)\Biggl)\alpha^i_{n+m-q}\alpha^i_q,\nn\\
	\eea
	where $[n]=k_1$ and $[m]=k_2$. The $\alpha^i_{n+m-q}\alpha^i_q$ part of commutator~\eqref{Zngeneralization1} comes from
	\bea
	[U^{k_1}_n,U^{k_2}_m]+[V^{k_1}_n,V^{k_2}_m].
	\eea
	Using the commutators~\eqref{vcommutator1} and \eqref{ucommutator1}, we should have the following identity
    \begin{widetext}
\bea
	&&-\frac{1}{4}\Biggl((m-q)h^{k_2}(q)\Big(h^{k_1}(q-m)+h^{k_1}(n+m-q)\Big)+(q-n)h^{k_2}(q-n)\Big(h^{k_1}(q)+h^{k_1}(n-q)\Big)\Biggl)\nn\\
	&&-\frac{1}{4}\Biggl((m-q)f^{k_2}(q)\Big(f^{k_1}(n+m-q)-f^{k_1}(q-m)\Big)-f^{k_2}(q-n)(q-n)\Big(f^{k_1}(q)-f^{k_1}(n-q)\Big)\Biggl)\nn\\
	&&=\frac{1}{2}h^{k_1+k_2}(q)(n-m)
\eea
\end{widetext}
	
	By substituting the expansions~\eqref{cons11} and~\eqref{cons21} into the commutator, after a tedious but straightforward calculation, one arrives at the following result:
	 \begin{widetext}
\bea
	\label{totallcommutator1}
	&&\frac{1}{2}\sum_q\sum_{w_1,w_2}\left[(n-q)\Big(d^{k_1}_{w_1}d^{k_2}_{w_2}-c^{k_1}_{w_1}c^{k_2}_{w_2}\Big)\zeta_{k_1}^{-w_2}+(q-m)\Big(d^{k_1}_{w_1}d^{k_2}_{w_2}-c^{k_1}_{w_1}c^{k_2}_{w_2}\Big)\zeta_{k_2}^{-w_1}\right]e^{\frac{2\pi i(w_1+w_2)}{N}q}\alpha^{i}_{n+m-q}\alpha^i_q\nn\\
	+&&
	\frac{1}{2}\sum_q\sum_{w_1,w_2}\left[(n-q)\Big(d^{k_1}_{w_1}c^{k_2}_{w_2}+c^{k_1}_{w_1}d^{k_2}_{w_2}\Big)\zeta_{k_1}^{-w_2}+(q-m)\Big(d^{k_1}_{w_1}c^{k_2}_{w_2}+c^{k_1}_{w_1}d^{k_2}_{w_2}\Big)\zeta_{k_2}^{-w_1}\right]e^{\frac{2\pi i(w_1+w_2)}{N}q}\epsilon^{ij}\alpha^{i}_{n+m-q}\alpha^j_q,\nn\\
	\eea
\end{widetext}

	A further simplification can be achieved by performing the change of summation variable $q \to n+m-q$ and applying the constraints~\eqref{cons11} and~\eqref{cons21} on the coefficients. 
	By including the $[U^{k_1}_n, V_m^{k_2}]+[V^{k_1}, U_m^{k_2}]$ part, this leads to the following consistency conditions for the closure of the algebra:
\begin{align}
&\sum_{q,w_1,w_2}(d^{k_1}_{w_1}c^{k_2}_{w_2}+c^{k_1}_{w_1}d^{k_2}_{w_2})(\zeta_{k_1}^{-w_2}-\zeta_{k_2}^{-w_1}) \nonumber\\
&\qquad\times e^{\frac{2\pi i(w_1+w_2)}{N}q}\epsilon^{ij}\alpha^i_{n+m-q}\alpha^j_q = 0, \\
&\sum_{q,w_1,w_2}(d^{k_1}_{w_1}d^{k_2}_{w_2}-c^{k_1}_{w_1}c^{k_2}_{w_2})(\zeta_{k_1}^{-w_2}-\zeta_{k_2}^{-w_1}) \nonumber\\
&\qquad\times e^{\frac{2\pi i(w_1+w_2)}{N}q}\alpha^i_{n+m-q}\alpha^i_q = 0,
\end{align}
It is crucial for the derivation of the above relation that the sum of the mode indices satisfies  $[n+m]=k_1+k_2$, ensuring the consistency of the mod $N$ structure in the algebra. 
	
	Using these simplifications, the net result of the commutator~\eqref{totallcommutator1} can be written as
		\begin{widetext}
\bea
	&&=\frac{1}{2}(n-m)\sum_{q,w_1,w_2}(d^{k_1}_{w_1}d^{k_2}_{w_2}-c^{k_1}_{w_1}c^{k_2}_{w_2})(\frac{\zeta_{k_1}^{-w_2}+\zeta_{k_2}^{-w_1}}{2})e^{\frac{2\pi i(w_1+w_2)}{N}q}\alpha^i_{n+m-q}\alpha^i_q\nn\\
	&&+\frac{1}{2}(n-m)\sum_{q,w_1,w_2}(d^{k_1}_{w_1}c^{k_2}_{w_2}+c^{k_1}_{w_1}d^{k_2}_{w_2})(\frac{\zeta_{k_1}^{-w_2}+\zeta_{k_2}^{-w_1}}{2})e^{\frac{2\pi i(w_1+w_2)}{N}q}\epsilon^{ij}\alpha^i_{n+m-q}\alpha^j_q\nn\\\label{qterm11}
	&&+\frac{1}{2}\sum_{q,w_1,w_2}(d^{k_1}_{w_1}d^{k_2}_{w_2}-c^{k_1}_{w_1}c^{k_2}_{w_2})(\zeta_{k_1}^{-w_2}-\zeta_{k_2}^{-w_1})e^{\frac{2\pi i(w_1+w_2)}{N}q}q\alpha^i_{n+m-q}\alpha^i_q\nn\\\label{qterm21}
	&&+\frac{1}{2}\sum_{q,w_1,w_2}(d^{k_1}_{w_1}c^{k_2}_{w_2}+c^{k_1}_{w_1}d^{k_2}_{w_2})(\zeta_{k_1}^{-w_2}-\zeta_{k_2}^{-w_1})e^{\frac{2\pi i(w_1+w_2)}{N}q}q\epsilon^{ij}\alpha^i_{n+m-q}\alpha^j_q
	\eea
\end{widetext}
	To be compatible with the commutator~\eqref{Zngeneralization1}, it is necessary that the $q\alpha^i_{n+m-q}\alpha^i_q$ and $q\epsilon^{ij}\alpha^i_{n+m-q}\alpha^j_q$ terms vanish, and the first two terms satisfy the following identities
	\bea
	\label{eq11}
	&&\sum_{\substack{w_1,w_2 \\ w_1+w_2=s~mod~N}}(d^{k_1}_{w_1}d^{k_2}_{w_2}-c^{k_1}_{w_1}c^{k_2}_{w_2})(\frac{\zeta_{k_1}^{-w_2}+\zeta_{k_2}^{-w_1}}{2})=d^{k_1+k_2}_s\nn\\\label{eq12}
	&&\sum_{\substack{w_1,w_2 \\ w_1+w_2=s~mod~N}}(d^{k_1}_{w_1}c^{k_2}_{w_2}+c^{k_1}_{w_1}d^{k_2}_{w_2})(\frac{\zeta_{k_1}^{-w_2}+\zeta_{k_2}^{-w_1}}{2})=c^{k_1+k_2}_s,\nn\\
	\eea
	where $s=0,1,...,N-1$ mod N, which are exactly the equations~\eqref{eq1}--\eqref{eq4}.
	
	\section{The $\mathbb{Z}_4$-equivariant Virasoro equations}
	\label{AppB}
	In this appendix, we present the equations~\eqref{eq1}--\eqref{eq4} explicitly, as they were manipulated using \texttt{Mathematica}. The equations have also been simplified by incorporating the constraint~\eqref{cons1}. 
	 \begin{widetext}

	 \begin{equation}
	 	\begin{aligned}
	 		&(c_2^1)^2 + (d_0^1)^2 = d_0^2 \\
	 		&(-1 - i) c_1^1 c_2^1 + (1 - i) d_0^1 d_1^1 = d_1^2 \\
	 		&(1 - i) c_1^1 d_0^1 - (1 + i) c_2^1 d_1^1 = c_1^2 \\
	 		&2i \left((c_1^1)^2 - (d_1^1)^2\right)= d_2^2 \\
	 		&(1 + i) c_1^1 c_2^1 - (1 - i) d_0^1 d_1^1 = -d_1^2 \\
	 		&(1 - i) c_1^1 d_0^1 - (1 + i) c_2^1 d_1^1 = c_1^2 \\
	 		&d_0^1 d_0^2 - \left(\frac{1}{2} - \frac{i}{2}\right) \left(-c_1^1 c_1^2 - d_1^1 d_1^2\right) - \left(\frac{1}{2} + \frac{i}{2}\right) \left(i c_1^1 c_1^2 + i d_1^1 d_1^2\right) = d_0^3 \\
	 		&\left(-\frac{1}{2} + \frac{i}{2}\right) \left(c_1^2 d_1^1 - c_1^1 d_1^2\right) - \left(\frac{1}{2} + \frac{i}{2}\right) \left(i c_1^2 d_1^1 - i c_1^1 d_1^2\right) = 0 \\
	 		&\left(-\frac{1}{2} - \frac{i}{2}\right) c_2^1 c_1^2 + \left(\frac{1}{2} - \frac{i}{2}\right) d_0^1 d_1^2 - i d_1^1 d_2^2 = d_1^3 \\
	 		&\left(\frac{1}{2} - \frac{i}{2}\right) c_1^2 d_0^1 - \left(\frac{1}{2} + \frac{i}{2}\right) c_2^1 d_1^2 + i c_1^1 d_2^2 = c_1^3 \\
	 		&\left(-\frac{1}{2} + \frac{i}{2}\right) \left(i c_1^1 c_1^2 - i d_1^1 d_1^2\right) - \left(\frac{1}{2} + \frac{i}{2}\right) \left(-c_1^1 c_1^2 + d_1^1 d_1^2\right) = 0 \\
	 		&c_2^1 d_0^2 - \left(\frac{1}{2} - \frac{i}{2}\right) \left(i c_1^2 d_1^1 + i c_1^1 d_1^2\right) - \left(\frac{1}{2} + \frac{i}{2}\right) \left(c_1^2 d_1^1 + c_1^1 d_1^2\right) = c_2^3 \\
	 		&\left(-\frac{1}{2} + \frac{i}{2}\right) c_2^1 c_1^2 - \left(\frac{1}{2} + \frac{i}{2}\right) d_0^1 d_1^2 - d_1^1 d_2^2 = -i d_1^3 \\
	 		&\left(\frac{1}{2} + \frac{i}{2}\right) c_1^2 d_0^1 + \left(\frac{1}{2} - \frac{i}{2}\right) c_2^1 d_1^2 - c_1^1 d_2^2 = i c_1^3 \\
	 		&c_2^1 c_2^3 + d_0^1 d_0^3 + i \left(-i c_1^1 c_1^3 - i d_1^1 d_1^3\right) - i \left(i c_1^1 c_1^3 + i d_1^1 d_1^3\right) = 1 \\
	 		&\left(\frac{1}{2} + \frac{i}{2}\right) c_2^1 c_1^3 + \left(\frac{1}{2} - \frac{i}{2}\right) c_1^1 c_2^3 + \left(\frac{1}{2} + \frac{i}{2}\right) d_1^1 d_0^3 + \left(\frac{1}{2} - \frac{i}{2}\right) d_0^1 d_1^3 = 0 \\
	 		&\left(\frac{1}{2} - \frac{i}{2}\right) c_1^3 d_0^1 + \left(\frac{1}{2} - \frac{i}{2}\right) c_2^3 d_1^1 + \left(\frac{1}{2} + \frac{i}{2}\right) c_1^1 d_0^3 + \left(\frac{1}{2} + \frac{i}{2}\right) c_2^1 d_1^3 = 0 \\
	 		&\left(\frac{1}{2} + \frac{i}{2}\right) c_2^1 c_1^3 + \left(\frac{1}{2} - \frac{i}{2}\right) c_1^1 c_2^3 + \left(\frac{1}{2} + \frac{i}{2}\right) d_1^1 d_0^3 + \left(\frac{1}{2} - \frac{i}{2}\right) d_0^1 d_1^3 = 0 \\
	 		&\left(-\frac{1}{2} + \frac{i}{2}\right) c_1^3 d_0^1 - \left(\frac{1}{2} - \frac{i}{2}\right) c_2^3 d_1^1 - \left(\frac{1}{2} + \frac{i}{2}\right) c_1^1 d_0^3 - \left(\frac{1}{2} + \frac{i}{2}\right) c_2^1 d_1^3 = 0 \\
	 		&2 (c_1^2)^2 + (d_0^2)^2 + 2 (d_1^2)^2 + (d_2^2)^2 = 1 \\
	 		&2 (c_1^2)^2 - 2 (d_1^2)^2 + 2 d_0^2 d_2^2 = 0 \\
	 		&d_0^2 d_0^3 - \left(\frac{1}{2} + \frac{i}{2}\right) \left(-c_1^2 c_1^3 - d_1^2 d_1^3\right) - \left(\frac{1}{2} - \frac{i}{2}\right) \left(-i c_1^2 c_1^3 - i d_1^2 d_1^3\right) = d_0^1 \\
	 				\end{aligned}
 		\end{equation}
 	\end{widetext}
    \begin{widetext}

	 		\begin{equation}
	 			\begin{aligned}
	 		&\left(-\frac{1}{2} + \frac{i}{2}\right) \left(i c_1^3 d_1^2 - i c_1^2 d_1^3\right) - \left(\frac{1}{2} + \frac{i}{2}\right) \left(-c_1^3 d_1^2 + c_1^2 d_1^3\right) = 0 \\
	 		&\left(-\frac{1}{2} + \frac{i}{2}\right) c_1^2 c_2^3 + \left(\frac{1}{2} + \frac{i}{2}\right) d_1^2 d_0^3 + i d_2^2 d_1^3 = d_1^1 \\
	 		&\left(-\frac{1}{2} + \frac{i}{2}\right) c_2^3 d_1^2 - i c_1^3 d_2^2 + \left(\frac{1}{2} + \frac{i}{2}\right) c_1^2 d_0^3 = c_1^1 \\
	 		&\left(-\frac{1}{2} - \frac{i}{2}\right) \left(-i c_1^2 c_1^3 + i d_1^2 d_1^3\right) - \left(\frac{1}{2} - \frac{i}{2}\right) \left(-c_1^2 c_1^3 + d_1^2 d_1^3\right) = 0 \\
	 		&c_2^3 d_0^2 - \left(\frac{1}{2} + \frac{i}{2}\right) \left(-i c_1^3 d_1^2 - i c_1^2 d_1^3\right) - \left(\frac{1}{2} - \frac{i}{2}\right) \left(c_1^3 d_1^2 + c_1^2 d_1^3\right) = c_2^1 \\
	 		&\left(-\frac{1}{2} - \frac{i}{2}\right) c_1^2 c_2^3 - \left(\frac{1}{2} - \frac{i}{2}\right) d_1^2 d_0^3 - d_2^2 d_1^3 = i d_1^1 \\
	 		&\left(\frac{1}{2} + \frac{i}{2}\right) c_2^3 d_1^2 - c_1^3 d_2^2 + \left(\frac{1}{2} - \frac{i}{2}\right) c_1^2 d_0^3 = -i c_1^1 \\
	 		&(c_2^3)^2 + (d_0^3)^2 = d_0^2 \\
	 		&(-1 + i) c_1^3 c_2^3 + (1 + i) d_0^3 d_1^3 = d_1^2 \\
	 		&(1 + i) c_1^3 d_0^3 - (1 - i) c_2^3 d_1^3 = c_1^2 \\
	 		&-i \left((c_1^3)^2 - (d_1^3)^2\right) + i \left(-(c_1^3)^2 + (d_1^3)^2\right) = d_2^2 \\
	 		&(1 - i) c_1^3 c_2^3 - (1 + i) d_0^3 d_1^3 = -d_1^2 \\
	 		&(1 + i) c_1^3 d_0^3 - (1 - i) c_2^3 d_1^3 = c_1^2\\
			&\left(\frac{1}{2}
            - \frac{i}{2}\right) c_2^1 c_1^2 + d_1^1 d_0^2 - \left(\frac{1}{2} + \frac{i}{2}\right) d_0^1 d_1^2 = 0 \\
			&\left(-\frac{1}{2} - \frac{i}{2}\right) c_1^2 d_0^1 + c_1^1 d_0^2 + \left(\frac{1}{2} - \frac{i}{2}\right) c_2^1 d_1^2 = 0 \\
			&\left(\frac{1}{2} + \frac{i}{2}\right) \left(i c_1^1 c_1^2 - i d_1^1 d_1^2\right) + \left(\frac{1}{2} - \frac{i}{2}\right) \left(-c_1^1 c_1^2 + d_1^1 d_1^2\right) - d_0^1 d_2^2 = 0 \\
			&\left(\frac{1}{2} + \frac{i}{2}\right) \left(i c_1^2 d_1^1 + i c_1^1 d_1^2\right) + \left(\frac{1}{2} - \frac{i}{2}\right) \left(c_1^2 d_1^1 + c_1^1 d_1^2\right) = 0 \\
			&\left(\frac{1}{2} + \frac{i}{2}\right) c_2^1 c_1^2 + i d_1^1 d_0^2 + \left(\frac{1}{2} - \frac{i}{2}\right) d_0^1 d_1^2 = 0 \\
			&\left(-\frac{1}{2} + \frac{i}{2}\right) c_1^2 d_0^1 - i c_1^1 d_0^2 - \left(\frac{1}{2} + \frac{i}{2}\right) c_2^1 d_1^2 = 0 \\
			&\left(\frac{1}{2} - \frac{i}{2}\right) c_2^1 c_1^3 - \left(\frac{1}{2} + \frac{i}{2}\right) c_1^1 c_2^3 + \left(\frac{1}{2} - \frac{i}{2}\right) d_1^1 d_0^3 - \left(\frac{1}{2} + \frac{i}{2}\right) d_0^1 d_1^3 = 0 \\
			&\left(-\frac{1}{2} - \frac{i}{2}\right) c_1^3 d_0^1 - \left(\frac{1}{2} + \frac{i}{2}\right) c_2^3 d_1^1 + \left(\frac{1}{2} - \frac{i}{2}\right) c_1^1 d_0^3 + \left(\frac{1}{2} - \frac{i}{2}\right) c_2^1 d_1^3 = 0 \\
			&-c_2^3 d_0^1 + c_2^1 d_0^3 + i \left(-c_1^3 d_1^1 - c_1^1 d_1^3\right) - i \left(c_1^3 d_1^1 + c_1^1 d_1^3\right) = 0 \\
				&\left(\frac{1}{2} + \frac{i}{2}\right) \left(c_1^2 d_1^1 - c_1^1 d_1^2\right) + \left(\frac{1}{2} - \frac{i}{2}\right) \left(i c_1^2 d_1^1 - i c_1^1 d_1^2\right) - c_2^1 d_2^2 = 0 \\
					&\left(\frac{1}{2} + \frac{i}{2}\right) \left(-c_1^1 c_1^2 - d_1^1 d_1^2\right) + \left(\frac{1}{2} - \frac{i}{2}\right) \left(i c_1^1 c_1^2 + i d_1^1 d_1^2\right) = 0 \\
						\end{aligned}
		\end{equation}
		\end{widetext}	
        \begin{widetext}

				\begin{equation}
				\begin{aligned}
			&\left(-\frac{1}{2} + \frac{i}{2}\right) c_2^1 c_1^3 + \left(\frac{1}{2} + \frac{i}{2}\right) c_1^1 c_2^3 - \left(\frac{1}{2} - \frac{i}{2}\right) d_1^1 d_0^3 + \left(\frac{1}{2} + \frac{i}{2}\right) d_0^1 d_1^3 = 0 \\
			&\left(-\frac{1}{2} - \frac{i}{2}\right) c_1^3 d_0^1 - \left(\frac{1}{2} + \frac{i}{2}\right) c_2^3 d_1^1 + \left(\frac{1}{2} - \frac{i}{2}\right) c_1^1 d_0^3 + \left(\frac{1}{2} - \frac{i}{2}\right) c_2^1 d_1^3 = 0 \\
			&\left(-\frac{1}{2} - \frac{i}{2}\right) \left(-c_1^1 c_1^2 - d_1^1 d_1^2\right) - \left(\frac{1}{2} - \frac{i}{2}\right) \left(i c_1^1 c_1^2 + i d_1^1 d_1^2\right) = 0 \\
			&\left(-\frac{1}{2} - \frac{i}{2}\right) \left(c_1^2 d_1^1 - c_1^1 d_1^2\right) - \left(\frac{1}{2} - \frac{i}{2}\right) \left(i c_1^2 d_1^1 - i c_1^1 d_1^2\right) + c_2^1 d_2^2 = 0 \\
			&\left(-\frac{1}{2} + \frac{i}{2}\right) c_2^1 c_1^2 - d_1^1 d_0^2 + \left(\frac{1}{2} + \frac{i}{2}\right) d_0^1 d_1^2 = 0 \\
			&\left(\frac{1}{2} + \frac{i}{2}\right) c_1^2 d_0^1 - c_1^1 d_0^2 - \left(\frac{1}{2} - \frac{i}{2}\right) c_2^1 d_1^2 = 0 \\
			&\left(-\frac{1}{2} - \frac{i}{2}\right) \left(i c_1^1 c_1^2 - i d_1^1 d_1^2\right) - \left(\frac{1}{2} - \frac{i}{2}\right) \left(-c_1^1 c_1^2 + d_1^1 d_1^2\right) + d_0^1 d_2^2 = 0 \\
			&\left(-\frac{1}{2} - \frac{i}{2}\right) \left(i c_1^2 d_1^1 + i c_1^1 d_1^2\right) - \left(\frac{1}{2} - \frac{i}{2}\right) \left(c_1^2 d_1^1 + c_1^1 d_1^2\right) = 0 \\
			&\left(-\frac{1}{2} - \frac{i}{2}\right) c_2^1 c_1^2 - i d_1^1 d_0^2 - \left(\frac{1}{2} - \frac{i}{2}\right) d_0^1 d_1^2 = 0 \\
			&\left(\frac{1}{2} - \frac{i}{2}\right) c_1^2 d_0^1 + i c_1^1 d_0^2 + \left(\frac{1}{2} + \frac{i}{2}\right) c_2^1 d_1^2 = 0 \\
					\end{aligned}
		\end{equation}
         \end{widetext}

         \begin{widetext}

			\begin{equation}
				\begin{aligned}	
			&\left(-\frac{1}{2} + \frac{i}{2}\right) \left(-c_1^2 c_1^3 - d_1^2 d_1^3\right) - \left(\frac{1}{2} + \frac{i}{2}\right) \left(-i c_1^2 c_1^3 - i d_1^2 d_1^3\right) = 0 \\
			&c_2^3 d_2^2 - \left(\frac{1}{2} + \frac{i}{2}\right) \left(i c_1^3 d_1^2 - i c_1^2 d_1^3\right) - \left(\frac{1}{2} - \frac{i}{2}\right) \left(-c_1^3 d_1^2 + c_1^2 d_1^3\right) = 0 \\
			&\left(-\frac{1}{2} - \frac{i}{2}\right) c_1^2 c_2^3 + \left(\frac{1}{2} - \frac{i}{2}\right) d_1^2 d_0^3 - d_0^2 d_1^3 = 0 \\
			&-c_1^3 d_0^2 - \left(\frac{1}{2} + \frac{i}{2}\right) c_2^3 d_1^2 + \left(\frac{1}{2} - \frac{i}{2}\right) c_1^2 d_0^3 = 0 \\
			&d_2^2 d_0^3 - \left(\frac{1}{2} - \frac{i}{2}\right) \left(-i c_1^2 c_1^3 + i d_1^2 d_1^3\right) - \left(\frac{1}{2} + \frac{i}{2}\right) \left(-c_1^2 c_1^3 + d_1^2 d_1^3\right) = 0 \\
			&\left(-\frac{1}{2} + \frac{i}{2}\right) \left(-i c_1^3 d_1^2 - i c_1^2 d_1^3\right) - \left(\frac{1}{2} + \frac{i}{2}\right) \left(c_1^3 d_1^2 + c_1^2 d_1^3\right) = 0 \\
			&\left(-\frac{1}{2} + \frac{i}{2}\right) c_1^2 c_2^3 - \left(\frac{1}{2} + \frac{i}{2}\right) d_1^2 d_0^3 + i d_0^2 d_1^3 = 0 \\
			&-i c_1^3 d_0^2 + \left(\frac{1}{2} - \frac{i}{2}\right) c_2^3 d_1^2 + \left(\frac{1}{2} + \frac{i}{2}\right) c_1^2 d_0^3 = 0 \\
			&\left(-\frac{1}{2} + \frac{i}{2}\right) c_2^1 c_1^3 + \left(\frac{1}{2} + \frac{i}{2}\right) c_1^1 c_2^3 - \left(\frac{1}{2} - \frac{i}{2}\right) d_1^1 d_0^3 + \left(\frac{1}{2} + \frac{i}{2}\right) d_0^1 d_1^3 = 0 \\
			&\left(\frac{1}{2} + \frac{i}{2}\right) c_1^3 d_0^1 + \left(\frac{1}{2} + \frac{i}{2}\right) c_2^3 d_1^1 - \left(\frac{1}{2} - \frac{i}{2}\right) c_1^1 d_0^3 - \left(\frac{1}{2} - \frac{i}{2}\right) c_2^1 d_1^3 = 0 \\
			&c_2^3 d_0^1 - c_2^1 d_0^3 - i \left(-c_1^3 d_1^1 - c_1^1 d_1^3\right) + i \left(c_1^3 d_1^1 + c_1^1 d_1^3\right) = 0 \\
			&\left(\frac{1}{2} - \frac{i}{2}\right) c_2^1 c_1^3 - \left(\frac{1}{2} + \frac{i}{2}\right) c_1^1 c_2^3 + \left(\frac{1}{2} - \frac{i}{2}\right) d_1^1 d_0^3 - \left(\frac{1}{2} + \frac{i}{2}\right) d_0^1 d_1^3 = 0 \\
			&\left(\frac{1}{2} + \frac{i}{2}\right) c_1^3 d_0^1 + \left(\frac{1}{2} + \frac{i}{2}\right) c_2^3 d_1^1 - \left(\frac{1}{2} - \frac{i}{2}\right) c_1^1 d_0^3 - \left(\frac{1}{2} - \frac{i}{2}\right) c_2^1 d_1^3 = 0 \\
			&\left(\frac{1}{2} - \frac{i}{2}\right) \left(-c_1^2 c_1^3 - d_1^2 d_1^3\right) + \left(\frac{1}{2} + \frac{i}{2}\right) \left(-i c_1^2 c_1^3 - i d_1^2 d_1^3\right) = 0 \\
			&-c_2^3 d_2^2 + \left(\frac{1}{2} + \frac{i}{2}\right) \left(i c_1^3 d_1^2 - i c_1^2 d_1^3\right) + \left(\frac{1}{2} - \frac{i}{2}\right) \left(-c_1^3 d_1^2 + c_1^2 d_1^3\right) = 0 \\
			&\left(\frac{1}{2} + \frac{i}{2}\right) c_1^2 c_2^3 - \left(\frac{1}{2} - \frac{i}{2}\right) d_1^2 d_0^3 + d_0^2 d_1^3 = 0 \\
			&c_1^3 d_0^2 + \left(\frac{1}{2} + \frac{i}{2}\right) c_2^3 d_1^2 - \left(\frac{1}{2} - \frac{i}{2}\right) c_1^2 d_0^3 = 0 \\
			&-d_2^2 d_0^3 + \left(\frac{1}{2} - \frac{i}{2}\right) \left(-i c_1^2 c_1^3 + i d_1^2 d_1^3\right) + \left(\frac{1}{2} + \frac{i}{2}\right) \left(-c_1^2 c_1^3 + d_1^2 d_1^3\right) = 0 \\
			&\left(\frac{1}{2} - \frac{i}{2}\right) \left(-i c_1^3 d_1^2 - i c_1^2 d_1^3\right) + \left(\frac{1}{2} + \frac{i}{2}\right) \left(c_1^3 d_1^2 + c_1^2 d_1^3\right) = 0 \\
			&\left(\frac{1}{2} - \frac{i}{2}\right) c_1^2 c_2^3 + \left(\frac{1}{2} + \frac{i}{2}\right) d_1^2 d_0^3 - i d_0^2 d_1^3 = 0 \\
			&i c_1^3 d_0^2 - \left(\frac{1}{2} - \frac{i}{2}\right) c_2^3 d_1^2 - \left(\frac{1}{2} + \frac{i}{2}\right) c_1^2 d_0^3 = 0
		\end{aligned}
	\end{equation}
	\end{widetext}

	\bibliographystyle{apsrev4-1}
	\bibliography{mybib}  %Produces the bibliography via BibTeX.
\end{document}